\pdfoutput=1 

\documentclass[a4paper,11pt]{article}
\pdfoutput=1 

\usepackage{jheppub} 

\usepackage[T1]{fontenc} 

\usepackage{amsmath}
\usepackage{amssymb}
\usepackage{amsfonts}
\usepackage{bm}
\usepackage{color}
\usepackage{float}

\title{\boldmath Practical quasi parton distribution functions}

\author[a]{Tomomi Ishikawa,}
\author[b, c, d]{Yan-Qing Ma,}
\author[e, f]{Jian-Wei Qiu}
\author[g, e]{and Shinsuke Yoshida
              \footnote{Present address:
	                Theoretical Division, Los Alamos National Laboratory,
			Los Alamos, NM 87545, USA.}}
\affiliation[a]{RIKEN BNL Research Center,\\
                Brookhaven National Laboratory, Upton, New York 11973, USA}
\affiliation[b]{School of Physics and State Key Laboratory
		of Nuclear Physics and Technology, Peking University,\\
 		Beijing 100871, China}
\affiliation[c]{Center for High Energy physics, Peking University,\\
 		Beijing 100871, China}		
\affiliation[d]{Collaborative Innovation Center of Quantum Matter,\\
 		Beijing 100871, China}
\affiliation[e]{Physics Department, Brookhaven National Laboratory,\\
                Upton, New York 11973, USA}
\affiliation[f]{C.N. Yang Institute for Theoretical Physics
                and Department of Physics and Astronomy,
		Stony Brook University,\\
 		Stony Brook, NY 11794-3840, USA}
\affiliation[g]{Theoretical Research Division, Nishina Center, RIKEN,\\
                Wako, Ibaraki, 351-0198, Japan}

\emailAdd{tomomi@quark.phy.bnl.gov}
\emailAdd{yqma@pku.edu.cn}
\emailAdd{jqiu@bnl.gov}
\emailAdd{yoshida@nt.sc.niigata-u.ac.jp}

\abstract{
A completely new strategy to calculate parton distoribution functions
on the lattice has recently been proposed.
In this method, lattice calculable observables, called quasi distributions,
are related to normal distributions.
The quasi distributions are known to contain power-law UV divergences
arise from a Wilson line in the non-local operator, while the normal
distributions only have logatithmic UV divergences.
We propose possible method to subtract the power divegence to make the matching
of the quasi with the normal distributions well-defined.
We also demonstrate the matching of the quasi quark distribution 
between continuum and lattice implementing the power divergence subtraction.
The matching calculations are carried out by one-loop perturbation.}


\begin{document} 

\rightline{\sffamily RBRC-1199}

\maketitle
\flushbottom

\section{Introduction}
\label{SEC:Introduction}

Quantum chromodynamics (QCD) is described by a simple Lagrangian,
but has remarkably wide coverage of scales from
its elementary degrees of freedom, quarks and gluons, to 
emergence of hadrons, such as pions and protons, as their composite states.
QCD makes our universe complex and rich in physical phenomena and enables
our intelligence to appear.
Parton distribution functions (PDFs) are important quantities in both
nuclear and elementary particle physics, where QCD plays a leading role.
In the high-energy scattering process,
one of the key concepts is the so-called
``QCD collinear factorization theorem''~\cite{Collins:1989gx, Collins:1981uw}:
Infrared (IR) collinear divergences can be absorbed into nonperturbative
distribution functions and thus 
the scattering cross sections can be factorized into perturbative
hard parts and nonperturbative parton distributions.
The PDFs are universal, which means once they are extracted from
some scattering processes, they can be used for other processes
to be calculated.
This fact gives QCD a huge predictive power.

The PDFs has been obtained by global QCD analyses, in which experimental
data are used being combined with the perturbative
calculation of hard parts and
with fitting assumptions for PDFs.
The attempts have been successful to some extent,
but there are still large uncertainties in many parameter regions.
While there have been future experiment plans to tackle extracting
the PDFs more accurately,
such as Electron Ion Collider (EIC)~\cite{Accardi:2012qut},
they may not cover whole range of Bjorken-$x$,
the longitudinal momentum fraction $x$ of a parton in the nucleon.
The lattice QCD could provide a framework to calculate the PDFs
nonperturbatively from the first principle, it has
an intrinsic difficulty, though:
Lattice QCD cannot handle ``real time''.
The quark distribution functions are written in operator description
in the light-cone coordinates as
\begin{eqnarray}
q(x,\mu)=\int\frac{d\xi^-}{4\pi}e^{-ix\xi^-P^+}
\langle{\cal N}(P)|\overline{\psi}(\xi^-)\gamma^+
\exp\left(-ig\int_0^{\xi^-}d\eta^-A^+(\eta^-)\right)\psi(0)|{\cal N}(P)\rangle,
\;\;\;\;
\label{EQ:normal-quark-PDFs}
\end{eqnarray}
where an operator in the hadronic matrix element with
nucleon momentum along the $z$-direction, $P=(P^0,0,0,P^z)$,
is stretched in $\xi^-$ direction, $\mu$ is the renormalization scale.
We here define $\xi^{\pm}=(t\pm z)/\sqrt{2}$ as the light-cone coordinate.
While it is time-dependent and therefore cannot be treated in the lattice
QCD directly, one way to avoid this inaccessibility is computing moments
of the distribution functions, instead, and recovering
the full distributions by inverse of the Mellin transformation.
Although several calculations of the moments have been carried out
~\cite{Dolgov:2002zm, Gockeler:2004wp},
it turns out that higher Mellin moments are hard to obtain and the attempts
in this direction have not been successful so far.

Recently, Ji introduced, so-called, ``quasi'' distributions~\cite{Ji:2013dva}:
\begin{eqnarray}
\widetilde{q}(\tilde{x},\mu, P_z)=\int\frac{dz}{4\pi}e^{-iz\tilde{x}P_z}
\langle{\cal N}(P_z)|\overline{\psi}(z)\gamma^3
\exp\left(-ig\int_0^zdz'A_3(z')\right)\psi(0)|{\cal N}(P_z)\rangle,
\;\;\;\;
\label{EQ:quasi-quark-PDFs}
\end{eqnarray}
for quark distributions,
where the non-local operator is stretched in $z$-direction,
purely spatial direction.
Ji's observation is that the normal distributions (\ref{EQ:normal-quark-PDFs})
can be recovered from the quasi distributions (\ref{EQ:quasi-quark-PDFs})
by taking the $P_z\rightarrow\infty$ limit,
\begin{eqnarray}
 \widetilde{q}(\tilde{x},\mu, P_z)\xrightarrow[P_z\rightarrow\infty]{}
  q(x,\mu),
\end{eqnarray}
because the PDFs are boost invariant.
While the quasi distributions do not contain ``real-time'' any more,
there is a trade-off: an infinite hadron momentum, which cannot be realized
in the actual lattice QCD simulations.
To make this accessible, Ji introduced the large momentum effective
theory and the quasi distribution with finite $P_z$,
which is calculable on the lattice, is matched to the one with infinite
$P_z$, therefore normal distributions, as:
\begin{eqnarray}
 \widetilde{q}(x,\Lambda, P_z)=
  Z\left(x,\frac{\Lambda}{P_z}, \frac{\mu}{P_z}\right)\otimes q(x,\mu)
  +{\cal O}\left(\frac{\Lambda_{\rm QCD}^2}{P_z^2},
	    \frac{M^2}{P_z^2}\right),
\end{eqnarray}
where $\otimes$ represents a convolution with respect to $x$
and $M$ is a nucleon mass.
As $P_z\rightarrow\infty$, the matching factor $Z$ goes to one and
${\cal O}(1/P_z^2)$ corrections are dropped off.
It is claimed~\cite{Xiong:2013bka} that since the difference
between quasi and normal distributions is just whether
the longitudinal momentum is finite or infinite,
they could have the common IR structure, which should not be
changed by moving from one frame to the other.
The matching factor $Z$, therefore, could be IR divergence free and
perturbatively calculable.

The fact that the matching factor in Ji's approach is IR-safe reminds us
of the QCD collinear factorization.
As mentioned earlier, 
in the high-energy scattering process with large momentum scale $Q$,
the scattering cross section can be factorized into hard parts and
nonperturbative functions such as PDFs and
fragmentation functions up to an uncertainty of
${\cal O}(\Lambda_{\rm QCD}^2/Q^2)$.
All collinear divergences are absorbed into the nonperturbative
functions, the remaining hard part is, therefore, IR-safe and
perturbatively calculable.
Inspired by Ji's idea and extending it to more familiar picture based on
the collinear factorization, two of authors of the present paper
introduced a concept
of the collinear factorization into the lattice calculable
parton distribution functions~\cite{Ma:2014jla, Ma:2014jga}.
In this approach, we start with finding ``lattice cross sections''
which can be factorized into hard parts and
targeted nonperturbative functions,
along with the analogy of the collinear factorization in the high-energy
scattering process.
This factorization is schematically expressed as
\begin{eqnarray}
 \widetilde{\sigma}(x,\widetilde{\mu}^2, P_z)=
  \sum_{\alpha=\{q,\overline{q}, g\}}H_{\alpha}
  \left(x,\frac{\widetilde{\mu}}{P_z}, \frac{\widetilde{\mu}}{\mu}\right)
  \otimes f_{\alpha}(x,\mu^2)
  +{\cal O}\left(\frac{\Lambda_{\rm QCD}^2}{\widetilde{\mu}^2}\right),
\label{EQ:lattice_crossections}
\end{eqnarray}
where the left-hand side is lattice calculable cross sections and it is
factorizable into hard parts $H_{\alpha}$ and nonperturbative functions
$f_{\alpha}$ in the right-hand side.
Depending on the cross section in the left-hand side, quark ($q$),
anti-quark ($\overline{q}$) and gluon ($g$) distributions could be involved.
In the analogy with usual scatterings, $\widetilde{\mu}$ and $P_z$
correspond to the momentum transfer (resolution) and
the collision energy (parameter), respectively.
As several kinds of high-energy scattering cross section give common
distribution functions (universality), we could design bunch of types of
lattice calculable cross sections to give desired nonperturbative functions.
From this view, the quasi distribution (\ref{EQ:quasi-quark-PDFs}),
introduced by Ji, is a special case of the lattice cross section and
the quasi quark distributions are factorized into hard parts and
normal distributions.

Several exploratory studies of the lattice computation for
the quasi distributions have been
carried out since Ji's original proposal of the method
~\cite{Lin:2014zya, Chen:2016utp, Alexandrou:2015rja}.
When the quasi distributions or lattice calculable cross sections are
computed on the lattice, the matching to continuum counterparts is necessary,
because their ultraviolet (UV) renormalizations are generally different
with each other.
One of the purpose of this paper is to demonstrate this matching
by one-loop perturbation, aiming at clarifying the matching strategy and
the future development for the nonperturbative method.

Another important aspect to be addressed in this program
is power-law UV divergences which quasi distributions contain.
The power divergence basically originates from a Wilson line operator in
the non-local operator.
On the other hand, in the normal distribution case,
the power divergence does not exist and there are only logarithmic UV
divergences, meaning the UV behavior is completely different between
normal and quasi distributions.
Because the left-hand sides of eqs.~(\ref{EQ:quasi-quark-PDFs}) and
(\ref{EQ:lattice_crossections}) are measured quantities, that is,
nonperturbative quantities, the treatment of the power divergence should be
nonperturbatively performed,
otherwise the perturbative expansion does not make any sense.
In this paper, we propose a subtraction method for the power divergences
so that the perturbative calculation of hard parts is justified.\\

The paper is organized as follows.
In section~\ref{SEC:global_picture} we present a global matching strategy
to extract PDFs from quasi PDFs concentrating on quark distribution functions.
The discussion here is based on the collinear factorization approach and
we clarify possible steps for the extraction and systematics entering
into this method through lattice QCD simulations.
The subtraction of the power-law UV divergence is discussed
in section~\ref{SEC:subtracted_quasi}.
We first show the existence of the linear divergence in the quasi quark
distributions through a simple one-loop perturbative calculation and
then introduce its subtraction scheme.
The technicalities necessary in the matching between continuum and lattice 
are presented in section~\ref{SEC:matching_cont_latt}, where we explain
the matching procedure and introduce UV cutoff scheme in two and three
dimension to regulate UV divergences.
In section~\ref{SEC:lattice_pt} some details in the lattice perturbation
including the mean-field improvement are explained and we give the one-loop
perturbative results of coefficients for the matching between continuum
and lattice using the naive lattice fermion and the plaquette gluon action
for simplicity.
A summary and outlook are given in section~\ref{SEC:summary}.
Appendices are devoted to Feynman rules for the lattice perturbation
(appendix~\ref{SEC:feynman_rules}) and actual expressions of the
one-loop corrections on the lattice for the matching
(appendix~\ref{SEC:one-loop_lattice}).


\section{Comments on extracting the parton distribution functions
from the lattice QCD}
\label{SEC:global_picture}

It is beneficial to see the basic strategy and problems
for extracting the PDFs from the lattice QCD simulation and
possible systematic uncertainties come in.

\subsection{Basic matching strategy}

In this subsection, we discuss possible strategy for calculating
the quark distributions on the lattice through the quasi distribution,
especially using the collinear factorization
approach~\cite{Ma:2014jla}.
The matching in the collinear factorization picture is expressed as
\begin{eqnarray}
 \widetilde{q}_{\rm LATT}(x, a^{-1}, P_z)
  &=&
  Z\left(x, aP_z, a\widetilde{\mu}\right)
  \otimes\widetilde{q}(x, \widetilde{\mu}, P_z)
  +{\cal O}_{\rm LATT}\left(\Lambda_{\rm QCD}^2a^2\right)\nonumber\\
 &=&
  Z\left(x, aP_z, a\widetilde{\mu}\right)
  \otimes
  H\left(x,\frac{P_z}{\widetilde{\mu}}, \frac{\mu}{\widetilde{\mu}}\right)
  \otimes q(x,\mu)\nonumber\\
 &&
  +{\cal O}_{\rm LATT}\left(\Lambda_{\rm QCD}^2a^2\right)
  +{\cal O}_{\rm TWIST}
  \left(\frac{\Lambda_{\rm QCD}^2}{\widetilde{\mu}^2}\right).
\label{EQ:whole_matching}
\end{eqnarray}
The basic work-flow could be:
\begin{enumerate}
 \item Generate lattice quasi distribution data
       $\widetilde{q}_{\rm LATT}(x, a^{-1}, P_z)$
       at several lattice spacings $a$.
       The lattice spacings should be reasonably small and the hadron
       momentum in $z$-direction, $P_z$, is expected to be sensibly large.
 \item Match perturbatively the lattice data to its continuum counter part
       $\widetilde{q}(x, \widetilde{\mu}, P_z)$
       at some matching scale $\widetilde{\mu}$.
       The matching scale could be around the inverse of the lattice spacings,
       $\widetilde{\mu}\sim a^{-1}$, to avoid large logarithms in the
       perturbative matching factor $ Z\left(x, aP_z, a\widetilde{\mu}\right)$.
       If this matching can be carried out nonperturbatively and
       the step scaling technique~\cite{Luscher:1991wu}
       can be employed,
       the matching scale could be set to be higher than $a^{-1}$,
       making uncertainties in the continuum side small.
 \item Take a continuum limit to eliminate lattice artifacts of
       ${\cal O}_{\rm LATT}\left(\Lambda_{\rm QCD}^2a^2\right)$,
       where we assume current standards for the lattice setting and then
       ${\cal O}_{\rm LATT}\left(\Lambda_{\rm QCD}a\right)$ error is absent.
       After this step, we obtain continuum value of
       the quasi distribution at a scale $\widetilde{\mu}$,
       which is independent of the lattice action we employ.
 \item Extract distribution functions $q(x,\mu)$ at a factorization scale
       $\mu\sim\widetilde{\mu}$
       by combining with perturbative hard part calculation
       $H\left(x,P_z/\widetilde{\mu}, \mu/\widetilde{\mu}\right)$.
       The best results would have an uncertainty of 
       ${\cal O}_{\rm TWIST}(\Lambda_{\rm QCD}^2/\widetilde{\mu}^2)$,
       unless higher twist effects are not taken into account.
       In this process, a condition
       $P_z\sim\widetilde{\mu}$ would also be important to make the
       perturbative uncertainty in the hard part small.
       This condition, in turn, leads to the fact that the finiteness of $P_z$
       affects to the higher twist uncertainty.
 \item The factorization scale $\mu$ is evolved to some reference scale
       $\mu_0$ by using DGLAP equation~\cite{Gribov:1972ri, Lipatov:1974qm,
       Altarelli:1977zs, Dokshitzer:1977sg},
       where we fit the obtained data $q(x,\mu)$ to parameters of
       distribution functions such as in the global QCD analysis
       (with experimental data).
       \footnote{In the DGLAP evolution of the factorization scale $\mu$,
       distributions of other flavors are generally mixed.}
       The existence of several values of $\mu\sim\widetilde{\mu}$ could
       discriminate higher twist effects
       ${\cal O}_{\rm TWIST}(\Lambda_{\rm QCD}^2/\widetilde{\mu}^2)$.
       When the higher twist effects is significant, the evolved distribution
       functions by the DGLAP equation
       could give different value depending on the initial scale
       $\mu\sim\widetilde{\mu}$.
       To provide several values of $\mu$ for the analysis,
       we need lattice calculations with largely separated several set
       of lattice spacings or performing the step scaling to obtain
       several different $\widetilde{\mu}$.
\end{enumerate}
If both the lattice artifacts and the higher twist effects are compromised
to exist, the step 1, 2 and 5 above could be simplified to skip our
effort as:
\begin{itemize}
 \item[1'] Generate lattice quasi distribution data
	   $\widetilde{q}_{\rm LATT}(x, a^{-1}, P_z)$
	   at a lattice spacing $a$,
 \item[2'] Match perturbatively or nonperturbatively
	   the lattice data to its continuum counter part
	   at a matching scale $\widetilde{\mu}=a^{-1}$,
 \item[5'] The scale $\mu$ is evolved to some reference scale $\mu_0$
	   by using the DGLAP equation,
	   where we fit the obtained data to parameters of
	   distribution functions.
 \end{itemize}
This leads a reduced relation:
\begin{eqnarray}
 \widetilde{q}_{\rm LATT}(x, a^{-1}, P_z)
 &=&
  Z\left(x, aP_z, 1\right)
  \otimes H\left(x,aP_z, a\mu\right)\otimes q(x,\mu)
  +{\cal O}\left(\Lambda_{\rm QCD}^2a^2\right).
\label{EQ:whole_matching_simple}
\end{eqnarray}
In this case, we always have ${\cal O}(\Lambda_{\rm QCD}^2a^2)$ uncertainties,
mixed effects from both lattice artifacts and higher twists.
The large $P_z\sim a^{-1}$ is still required to make
perturbative uncertainty from the hard part small.

\subsection{IR consistency}

It has been shown that the quasi quark distributions have the same IR structure
as that in the normal quark distributions~\cite{Xiong:2013bka, Ma:2014jla}.
All the soft IR divergences cancel in both normal and quasi distribution.
While the collinear divergences do not vanish, they are the same between
the two distributions, meaning the quasi quark distributions
can be factorized into perturbative hard parts and nonperturbative
quark distributions.

The lattice theories are designed to share the same IR behavior with
its continuum counterpart and their difference is only in UV region,
meaning the lattice quasi distributions can also be factorized
into the hard part and the continuum quasi distributions.

\subsection{UV inconsistency}

As originally argued in ref.~\cite{Ji:2013dva}, quasi distributions
are power UV divergent,
while normal distributions are only logarithmically UV divergent.
To compensate these differences, the perturbative hard part must include
these UV divergences.
Although the notorious power-law UV divergence does not arise
in the dimensional regularization as in the $\overline{\rm MS}$ scheme,
we have to use lattice regularization,
which corresponds to UV cutoff regularization,
to extract the quasi distributions, where the power UV divergence is
manifest.
The power divergences can ruin the perturbative accuracy.
From the viewpoint of lattice QCD, there is no continuum limit
when the matching is perturbatively carried out.
To deal with the power divergence, its subtraction is essential.
The subtraction should be nonperturbative, otherwise, again,
cutoff of the perturbative expansion at some order could ruin the
perturbative accuracy.
The nonperturbative subtraction method for the power divergence is to be
discussed in the next section.


\section{Power divergence subtracted quasi distributions}
\label{SEC:subtracted_quasi}

The quasi PDFs are Fourier
transform of non-local operators which include a Wilson line operator,
where it is well-known that the Wilson lines generate notorious
``power-law UV divergences''.
The subtraction of the power divergences is essential to extract
physically meaningful results and it must be carried out nonperturbatively.
Although the subtraction of the divergences has been discussed
frequently since old days,
it would be beneficial to repeat the logic in this section.

\subsection{Quasi quark distribution functions}

We here again write the quasi quark distribution function denoted by
eq.~(\ref{EQ:quasi-quark-PDFs}):
\begin{eqnarray}
\widetilde{q}(\tilde{x}, P_z)
=\int\frac{d\delta z}{2\pi}e^{-i\tilde{x}P_z\delta z}
\langle{\cal N}(P_z)|O_{\delta z}|{\cal N}(P_z)\rangle,
\end{eqnarray}
where the non-local operator elongated in $z$-direction is defined as
\begin{eqnarray}
O_{\delta z}=
\int_x\overline{\psi}(x+\hat{\bm 3}\delta z)\gamma_3
U_3(x+\hat{\bm 3}\delta z; x)\psi(x),
\label{EQ:non-local_operator}
\end{eqnarray}
with a straight Wilson line operator
\begin{eqnarray}
 U_3(x\pm\hat{\bm 3}|\delta z|; x)=P
  \exp\left(-ig\int_0^{|\delta z|}dz'A_3(x\pm\hat{\bm 3}z')\right).
\end{eqnarray}
where $P$ denotes the path-ordering symbol meaning increasing $z'$
from right to left.
The essential part of the quasi distribution is a nonperturbative
matrix element, which we call non-local matrix element,
and we define
\begin{eqnarray}
{\cal M}_{\delta z}(P_z)=\langle{\cal N}(P_z)|O_{\delta z}|{\cal N}(P_z)\rangle.
\end{eqnarray}
In the matrix element, nucleon states have momentum in $z$-direction, $P_z$,
which we write as $|{\cal N}(P_z)\rangle$.
This matrix element is calculable using the lattice QCD simulation
since it is not time-dependent.
In the following, our calculations are performed in Euclidean space.

\subsection{First look at the one-loop amplitude of the non-local
   matrix element}
\label{SEC:first_look}

In this subsection, we take a glance at the structure of the non-local
matrix element ${\cal M}_{\delta z}(P_z)$
up to one-loop order to clarify the UV and IR divergence contained in it.

We first present Feynman rules for the non-local operator $O_{\delta z}$
(eq.~(\ref{EQ:non-local_operator})) in covariant gauge.
The rules up to ${\cal O}(g^2)$ are
\begin{eqnarray}
O_{\delta z}^{(0)}(p, q)
&=&
\gamma_3\delta(p-q)e^{-ip_3\delta z},
\label{EQ:feynman_rule_non-local-0}\\
O_{\delta z}^{(1)\mu, A}(p, q, k)
&=&
igT^A\gamma_3\delta^{\mu 3}\delta(k-p+q)e^{-ip_3\delta z}
\frac{1-e^{ik_3\delta z}}{ik_3},
\label{EQ:feynman_rule_non-local-1}\\
O_{\delta z}^{(2)\mu\nu, AB}(p, q, k)
&=&
-g^2\{T^A,T^B\}\gamma_3\delta^{\mu 3}\delta^{\nu 3}\delta(p-q)e^{-ip_3\delta z}
\left(\frac{1-e^{ik_3\delta z}}{k_3^2}-\frac{\delta z}{ik_3}\right),
\label{EQ:feynman_rule_non-local-2}
\end{eqnarray}
where their diagrammatic expressions are shown
in figure~\ref{FIG:feynman_rule_non-local}.
(See appendix \ref{APPENDIX:feynman_rules_non-local_operator} for the
derivation.)
We note that there is no IR pole structure
in eqs.~(\ref{EQ:feynman_rule_non-local-1})
and (\ref{EQ:feynman_rule_non-local-2}),
but a term proportional to $\delta z$
in eq.~(\ref{EQ:feynman_rule_non-local-2}) brings us a UV linear divergence,
which is to be shown soon.
When the seemingly convenient axial gauge ($A_3(x)=0$) is taken,
$O_{\delta z}^{(1)\mu, A}(p, q, k)$ and
$O_{\delta z}^{(2)\mu\nu, AB}(p, q, k)$ vanish,
making the Feynman rules for the non-local operator simple.
However we would have a drawback:
complications come into gluon propagators.
While the gluon propagator with the axial gauge has the similar structure
appears in $O_{\delta z}^{(1)\mu, A}(p, q, k)$ and
$O_{\delta z}^{(2)\mu\nu, AB}(p, q, k)$
above, there is spurious IR poles and then some pole prescription
is required.
We concentrate on the covariant gauge through out this paper.
\begin{figure}
\begin{center}
\parbox{45mm}{
\begin{center}
\includegraphics[scale=0.3, viewport = 0 0 350 330, clip]
{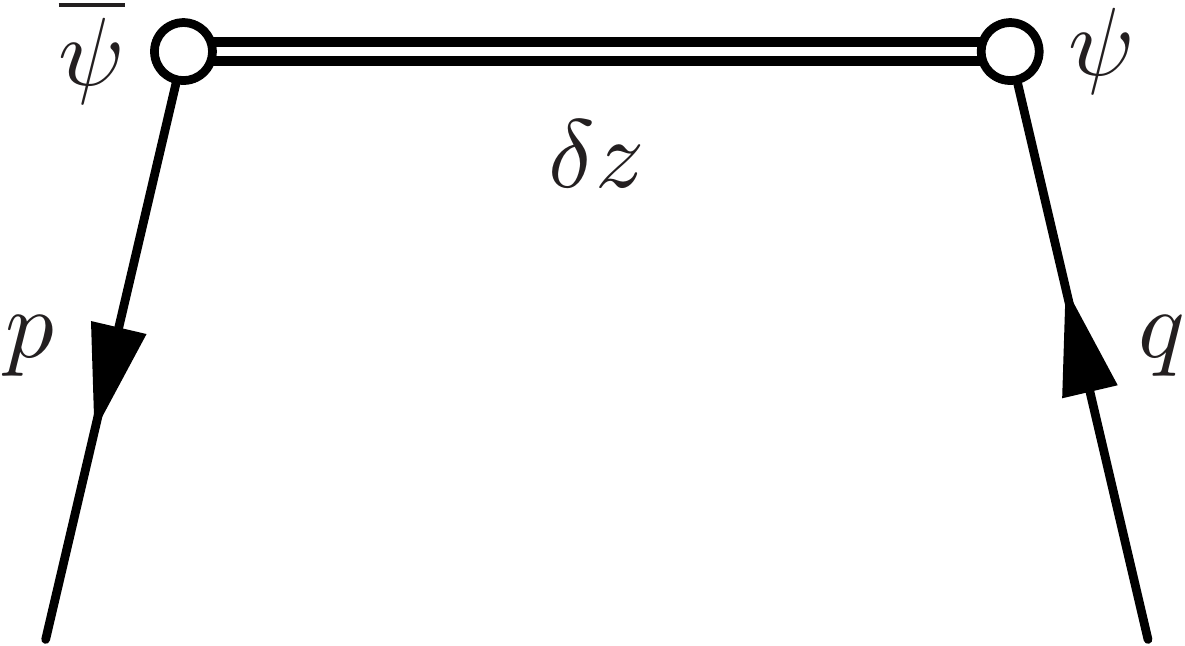}
$O_{\delta z}^{(0)}(p,q)$
\end{center}
}
\parbox{45mm}{
\begin{center} 
\includegraphics[scale=0.3, viewport = 0 0 350 330, clip]
{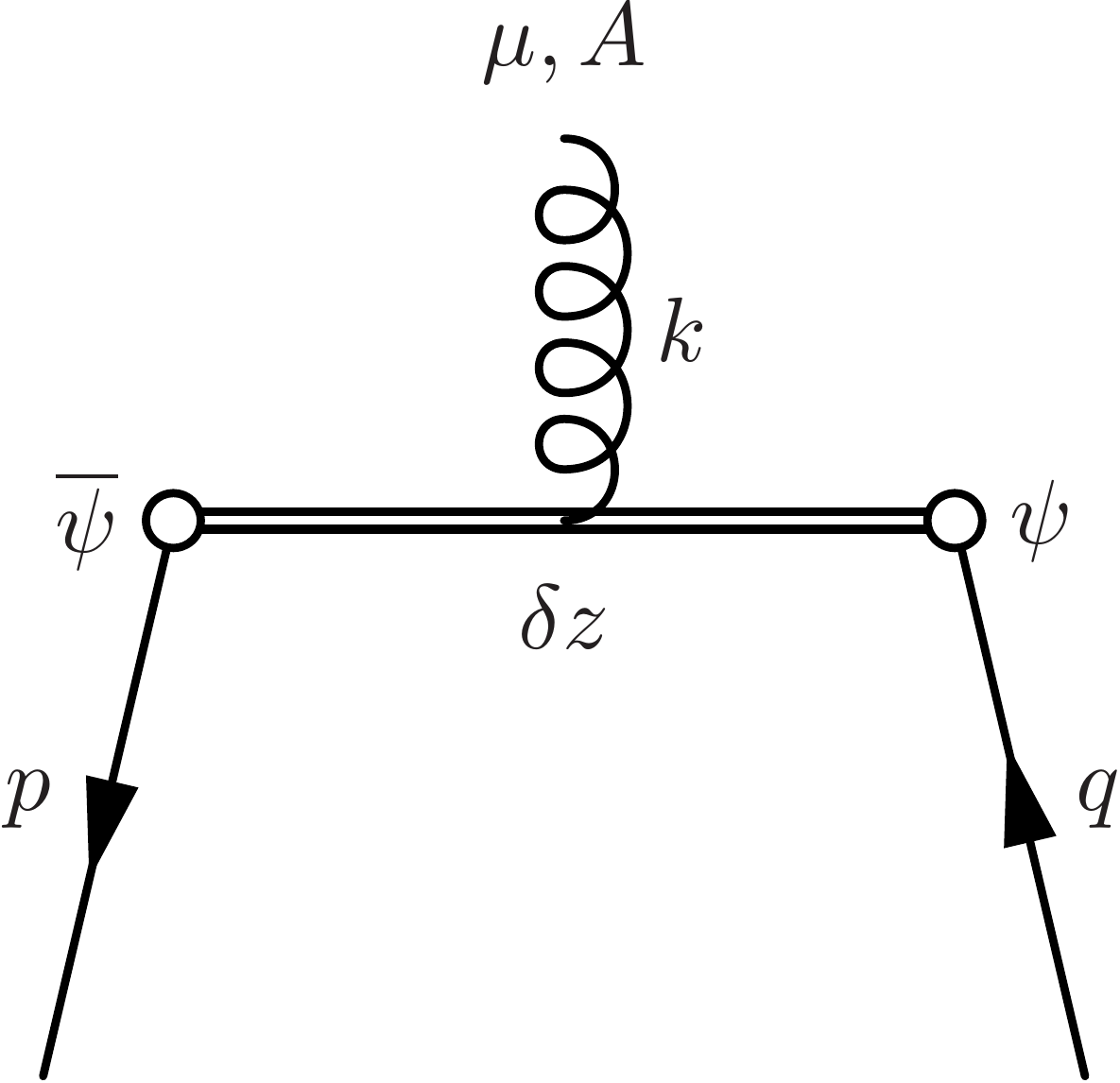} 
$O_{\delta z}^{(1)\mu, A}(p,q,k)$
\end{center}
}
\parbox{45mm}{
\begin{center}
\includegraphics[scale=0.3, viewport = 0 0 350 330, clip]
{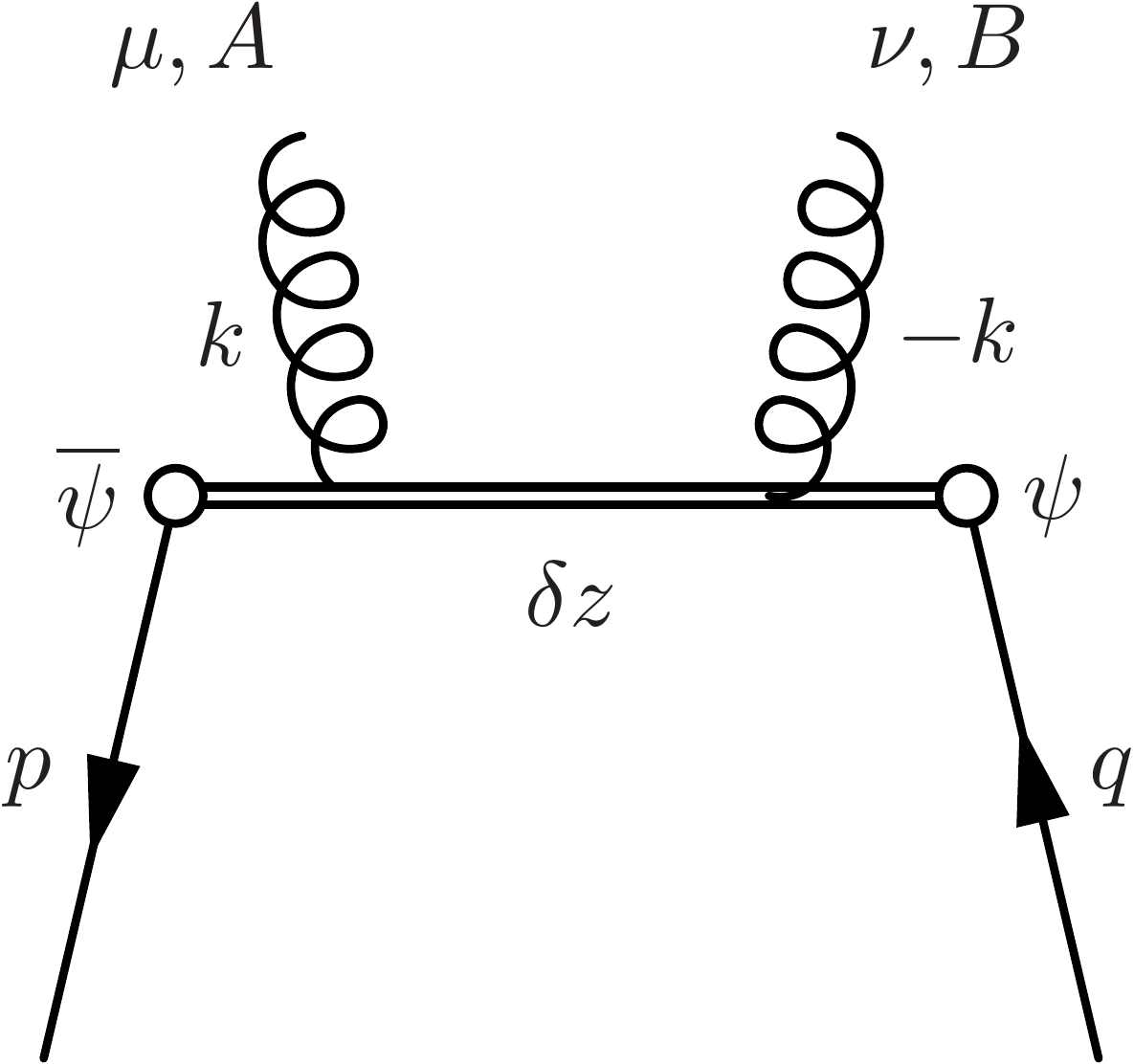}
$O_{\delta z}^{(2)\mu\nu, AB}(p,q,k)$
\end{center}
}
 \caption{Diagrammatic expression of Feynman rules for non-local
 quark bilinear operator relevant to one-loop perturbative calculation.
 For $O_{\delta z}^{(2)\mu\nu, AB}(p,q,k)$ (right),
 two gluon lines have a common momentum $k$ with opposite sign,
 where this setting is sufficient for the perturbative calculation
 at one-loop level.}
 \label{FIG:feynman_rule_non-local}
 \end{center}
\end{figure}

We consider a case with zero external momentum, $P_3=0$, and take 
the Feynman gauge ($\alpha=1$) for simplicity.
(In this paper, the matching between continuum
and lattice is carried out before $\delta z$ is integrated.
The introduction of the external momentum would not be necessary
in this matching, because they would share the same IR structure.)
At one-loop level in this gauge, there are three types of diagrams
to be calculated (vertex, sail and operator tadpole-type) shown
in figure~\ref{FIG:feynman_diagram_deltaGamma},
whose contributions to the one-loop amplitude are depicted as:
\begin{figure}
\begin{center}
\parbox{35mm}{
\begin{center}
\includegraphics[scale=0.3, viewport = 0 0 320 300, clip]
{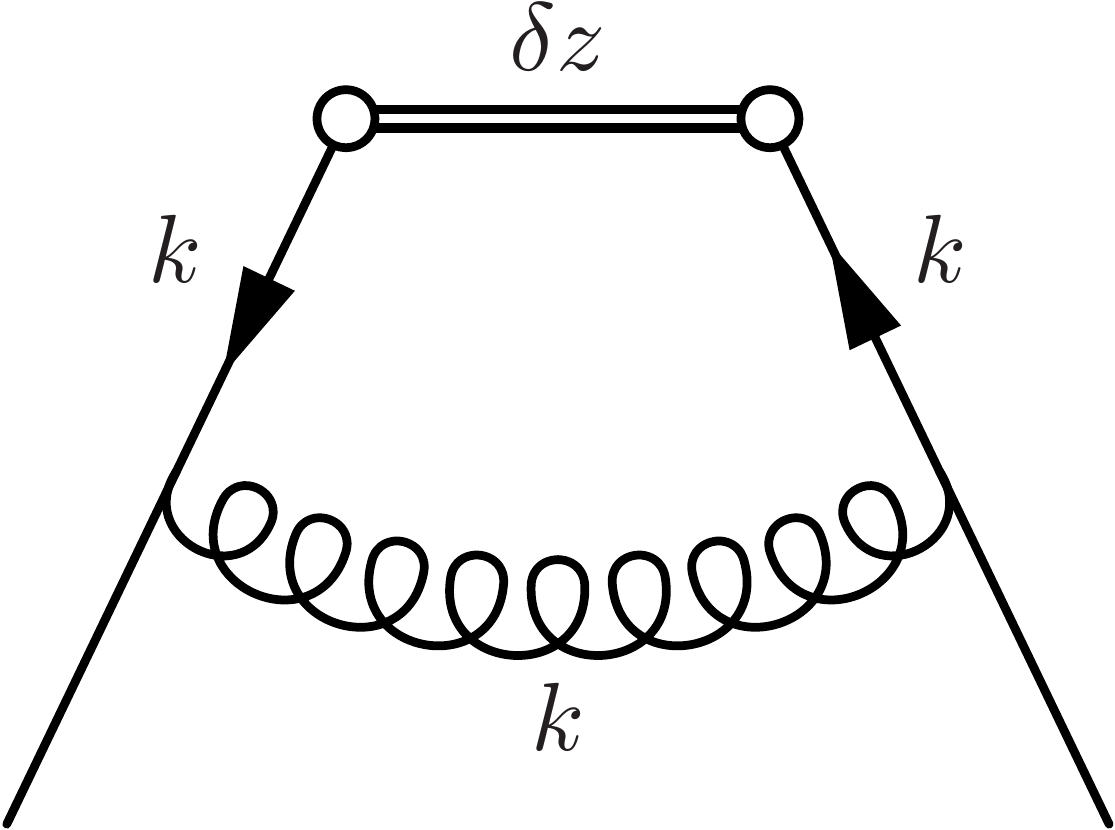}
$\delta\Gamma_{\rm vertex}(\delta z)$
\end{center}
}
\parbox{78mm}{
\begin{center}
\includegraphics[scale=0.3, viewport = 0 0 320 300, clip]
{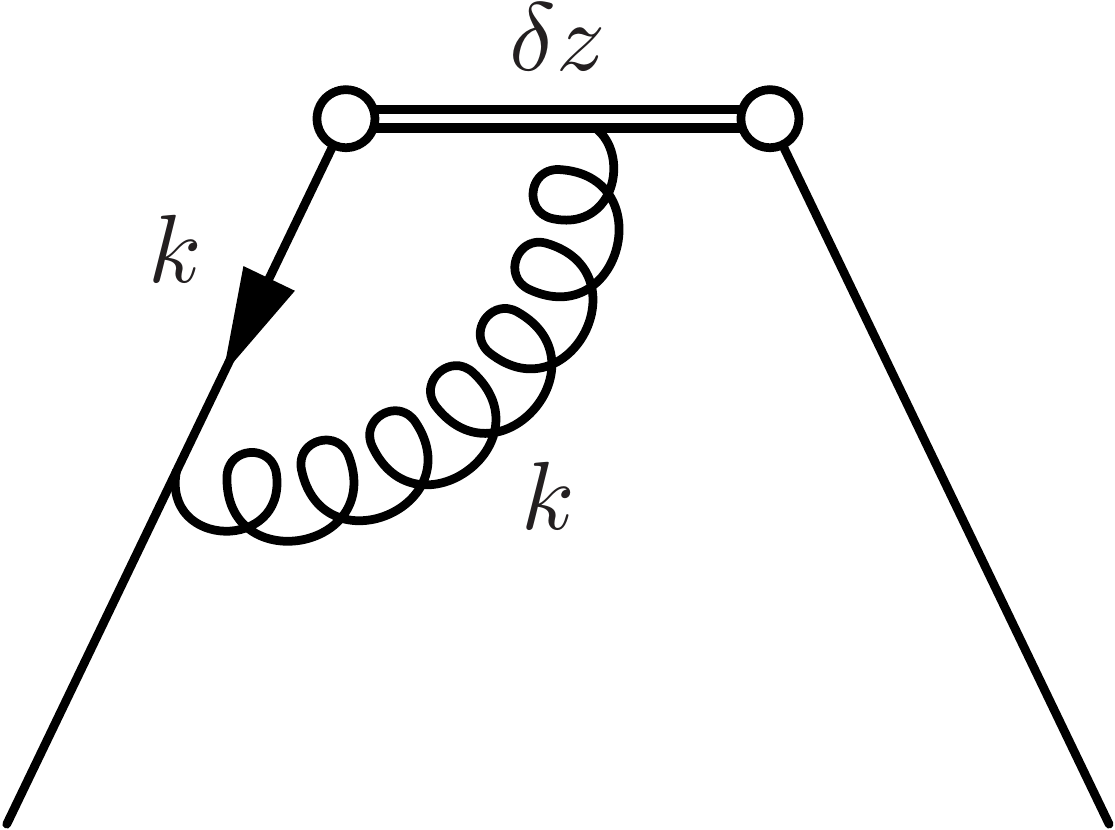}
\includegraphics[scale=0.3, viewport = 0 0 320 300, clip]
{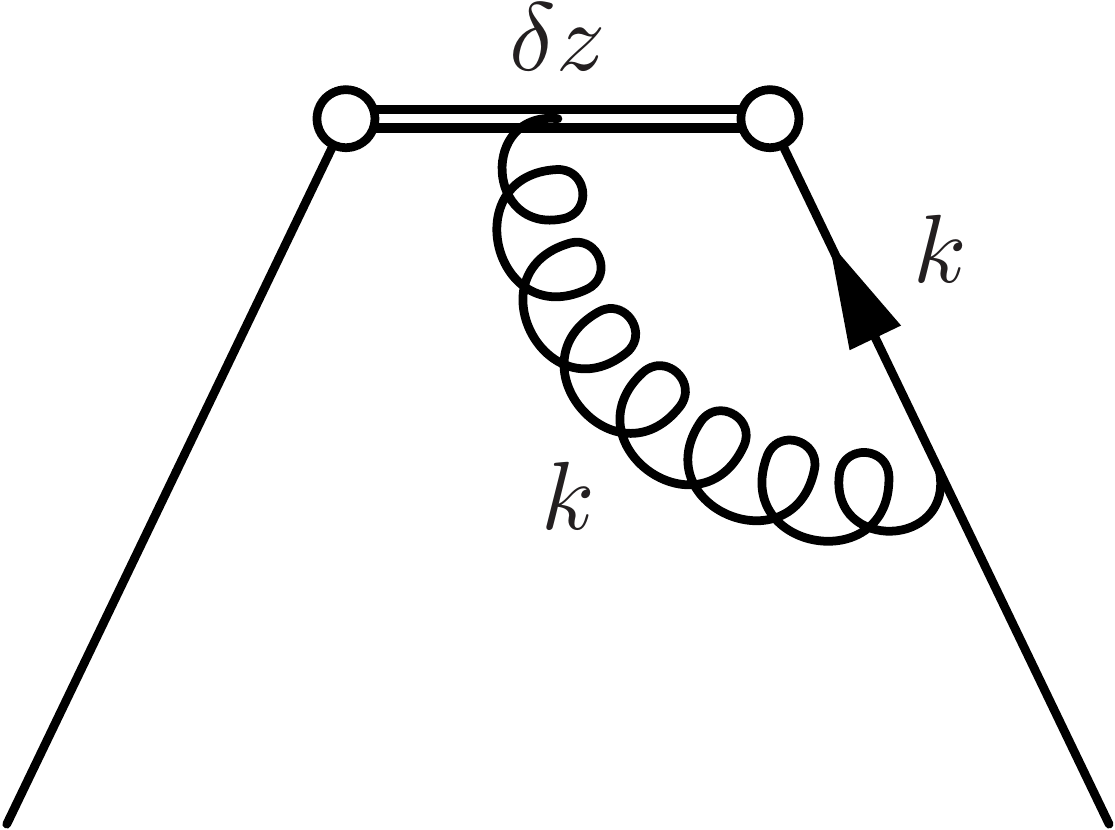} 
$\delta\Gamma_{\rm sail}(\delta z)$
\end{center}
}
\parbox{35mm}{
\begin{center} 
\includegraphics[scale=0.3, viewport = 0 0 320 320, clip]
{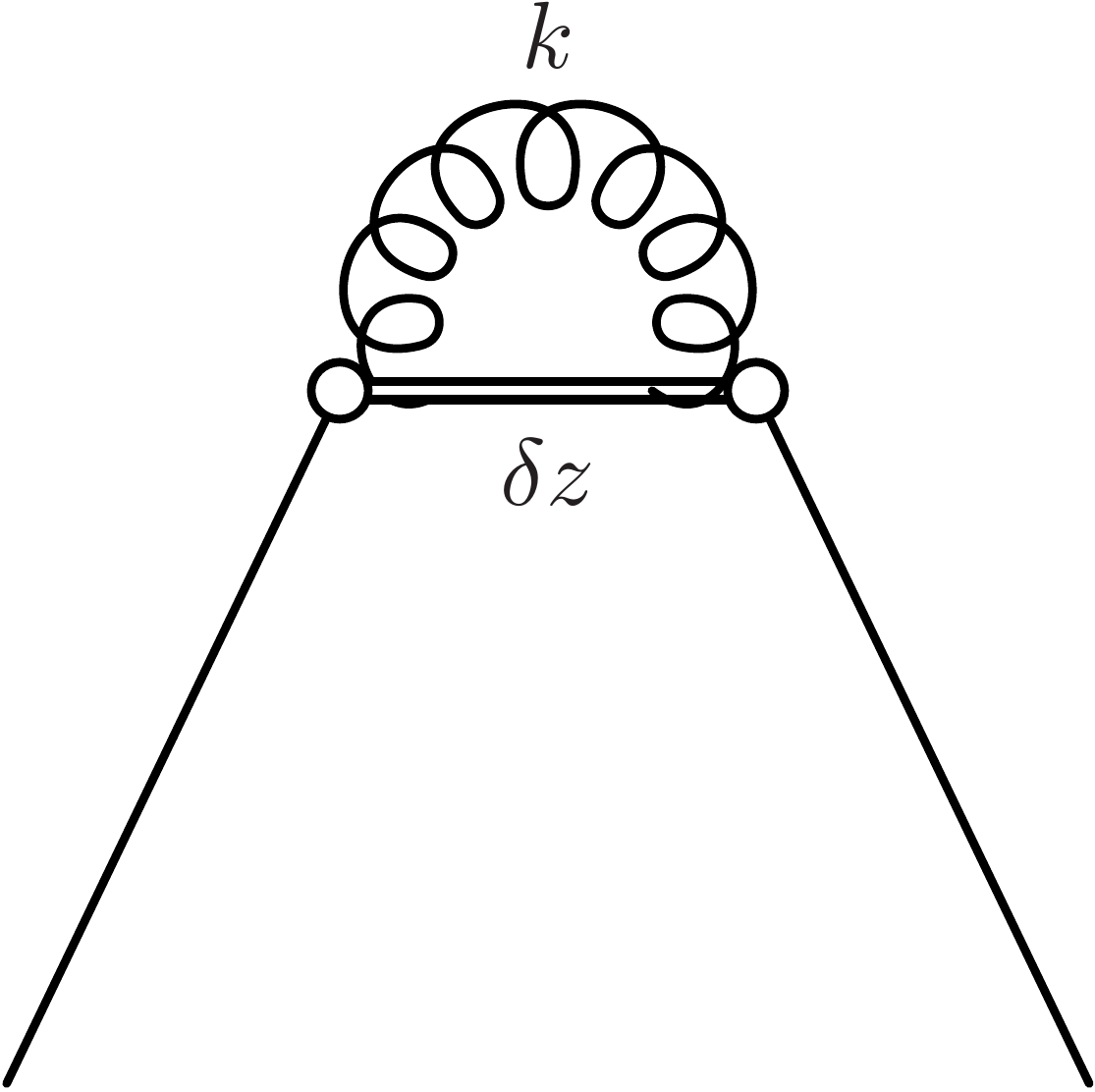} 
$\delta\Gamma_{\rm tadpole}(\delta z)$
\end{center}
}
\caption{One-loop diagrams for $\delta\Gamma$s.}
 \label{FIG:feynman_diagram_deltaGamma} 
 \end{center}
\end{figure}
 \begin{figure}
\begin{center}
\parbox{35mm}{
\includegraphics[scale=0.3, viewport = 0 0 320 170, clip]
{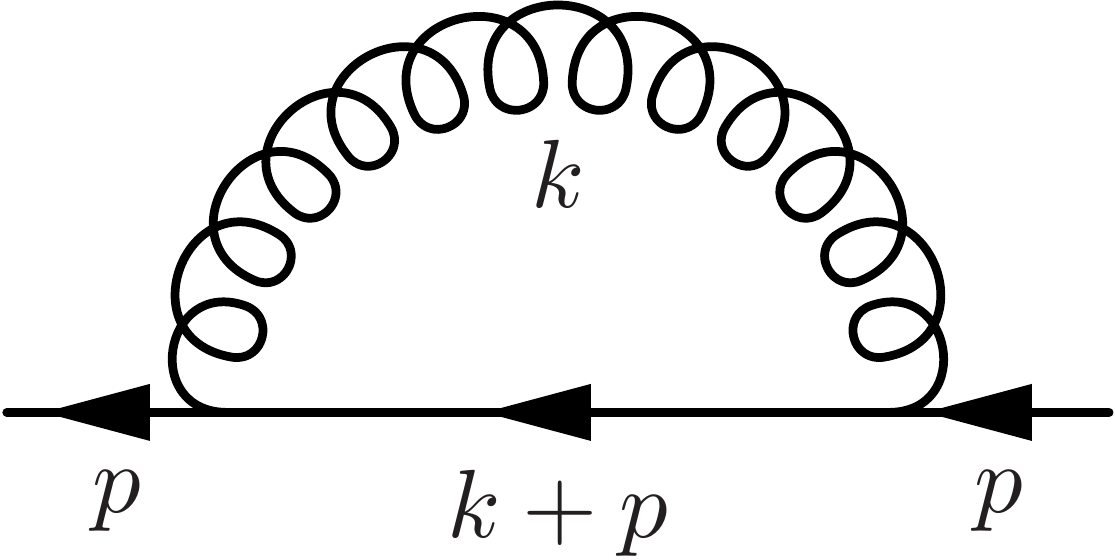}
}
\;\;\;$\Sigma_{\rm sunset}(p)$
\caption{Sunset diagram for quark wave-function renormalization.}
\label{FIG:feynman_diagram_sunset}
\end{center}  
\end{figure}
\begin{eqnarray}
\delta\Gamma_{\rm vertex}(\delta z)&=&
\int_kG_{\mu\nu}^{AB}(k)V_{\nu}^B(0,k)S_q(k)
O_{\delta z}^{(0)}(k, k)S_q(k)V_{\mu}^A(k,0)
\nonumber\\
&\equiv&\left(\frac{g}{4\pi}\right)^2C_F\gamma_3I_{\rm vertex}(\delta z),
\\
\delta\Gamma_{\rm sail}(\delta z)&=&
\int_k\left\{G_{\mu 3}^{AB}(k)O_{\delta z}^{(1)3, B}(0,k)S_q(k)V_{\mu}^A(k,0)
+V_{\mu}^A(0,k)S_q(k)O_{\delta z}^{(1)3, B}(k,0)G_{3\mu}^{BA}(k)\right\}
\nonumber\\
&\equiv&\left(\frac{g}{4\pi}\right)^2C_F\gamma_3I_{\rm sail}(\delta z),
\\
\delta\Gamma_{\rm tadpole}(\delta z)&=&
\int_kG_{33}^{AB}(k)O_{\delta z}^{(2)33, BA}(0,0,k)
\nonumber\\
&\equiv&\left(\frac{g}{4\pi}\right)^2C_F\gamma_3I_{\rm tadpole}(\delta z),
\end{eqnarray}
where $G_{\mu\nu}^{AB}(k)$, $S_q(k)$ and $V_{\mu}^A(k,l)$ are gluon propagator,
quark propagator and quark-gluon vertex, respectively, whose definitions
are seen in appendix~\ref{SEC:feynman_rules}.
$C_F$ is
the quadratic Casimir operator of the fundamental representation of
$SU(N_c)$, which is $C_F=(N_c^2-1)/(2N_c)$.
We first integrate out a loop-momentum $k_3$ introducing IR cutoff on
the loop momentum in perpendicular directions to $z$, $k_{\perp z}$, and obtain:
\begin{eqnarray}
I_{\rm vertex}(\delta z)&=&
(4\pi)^2\frac{d-2}{8}\int_{k_{\perp z}}
\left(\frac{1}{k_{\perp z}^3}+\frac{|\delta z|}{k_{\perp z}^2}
+\frac{|\delta z|^2}{k_{\perp z}}\right)e^{-k_{\perp z}|\delta z|},
\\
I_{\rm sail}(\delta z)&=&
(4\pi)^2\frac{1}{2}\int_{k_{\perp z}}
\left[\frac{1}{k_{\perp z}^3}
-\left(\frac{1}{k_{\perp z}^3}+\frac{|\delta z|}{k_{\perp z}^2}\right)
e^{-k_{\perp z}|\delta z|}\right],
\\
I_{\rm tadpole}(\delta z)&=&
(4\pi)^2\frac{1}{2}\int_{k_{\perp z}}
\left[\frac{1}{k_{\perp z}^3}-\frac{|\delta z|}{k_{\perp z}^2}
-\frac{1}{k_{\perp z}^3}e^{-k_{\perp z}|\delta z|}\right],
\end{eqnarray}
where $d$ is the dimension of Euclidean space and eventually
is set to be $d=4$.
Wave function renormalizations are also required,
it is, however, not different from the usual case, in which
there is not $\delta z$ dependence, and obtained just by calculating
quark self-energy $\Sigma_{\rm sunset}(p)$ from a sunset diagram
(figure~\ref{FIG:feynman_diagram_sunset}):
\begin{eqnarray}
Z_{\psi}=
1-\left.\frac{\partial\Sigma_{\rm sunset}(p)}{\partial i\!\!\not{\!p}}\right|_{p=0}
\equiv 1+\left(\frac{g}{4\pi}\right)^2C_F F+{\cal O}(g^4).
\end{eqnarray}
We make up a total one-loop amplitude,
\begin{eqnarray}
{\cal M}_{\delta z}(P_3=0)
=
\left[1+\left(\frac{g}{4\pi}\right)^2C_F{\cal A}(\delta z)+{\cal O}(g^4)\right]
\left[{\cal M}_{\delta z}(P_3=0)\right]^{\rm tree},
\label{EQ:one-loop_amplitude}
\end{eqnarray}
where $\left[{\cal M}_{\delta z}(P_3)\right]^{\rm tree}$
denotes the tree-level amplitude and
\begin{eqnarray}
{\cal A}(\delta z)=
I_{\rm vertex}(\delta z)+I_{\rm sail}(\delta z)+I_{\rm tadpole}(\delta z)+F.
\end{eqnarray}

When $\delta z=0$, a local operator case, contributions from
the sail- and tadpole-type diagrams vanish,
and $\delta\Gamma_{\rm vertex}$ gives logarithmic UV and IR divergences
in $d=4$ as in usual situation.
In the non-local case, $\delta\Gamma_{\rm vertex}$ is UV finite,
because the loop integral is regulated by $\delta z\not=0$,
whilst $\delta\Gamma_{\rm sail}$ and $\delta\Gamma_{\rm tadpole}$
are not finite in $d=4$.
Especially, $\delta\Gamma_{\rm tadpole}$ has a linear UV divergence which
originates from a Wilson line in the non-local operator.
More specifically, this power-law divergence structure is generated
from the second term in eq.~(\ref{EQ:feynman_rule_non-local-2}),
which is proportional to $\delta z$.
This linear UV divergence should be subtracted nonperturbatively
in the renormalization.
A perturbative subtraction makes no sense, because the observed
non-local matrix element itself is nonperturbative and the truncation
at some order in the coupling expansion would not absorb correctly
the UV inconsistency between normal and quasi distributions.
One viable option for the nonperturbative subtraction is to use a static
$q\bar{q}$ potential as explained in the next subsection.

\subsection{Renormalization of Wilson line and subtraction of
            linear UV divergence}

In order to subtract the linear divergence caused by the Wilson line,
a counter term needs to be introduced.
To prepare the counter term to all-order in $g$ systematically,
we use an observable which share the same power-law divergence structure
as in the non-local operator.
One of natural choices for the observable is a static $q\bar{q}$ potential,
in which the power divergence also originates from Wilson lines.

The renormalization of the non-local operator has long been known since
the 1980s~\cite{Dotsenko:1979wb, Arefeva:1980zd, Craigie:1980qs, Dorn:1986dt}
and has also been discussed in the context of static heavy quarks,
where the static heavy quark propagator is represented by
a straight Wilson line in temporal direction
~\cite{Boucaud:1989ga, Eichten:1989kb, Maiani:1991az, Martinelli:1998vt}.
(The renormalization of the non-local quark bilinear operator
is also well summarized in ref.~\cite{Musch:2010ka}.)
We here quote their conclusions.
A Wilson line along a (smooth) contour ${\cal C}$,
$W_{\cal C}$, is renormalized as
\begin{eqnarray}
 W_{\cal C}=Z_ze^{\delta m \ell({\cal C})}W_{\cal C}^{\rm ren},
\label{EQ:wilson-line_renormalization}
\end{eqnarray}
where a superscript ``ren'' indicates the operator is renormalized,
$\ell({\cal C})$ denotes length along the contour ${\cal C}$,
$\delta m$ is mass renormalization of a test particle moving along the contour
${\cal C}$.
The renormalization factor $Z_z$ arises from end points of the Wilson line,
where the subscript ``$z$'' indicates an auxiliary field
($z$-field) to represent a Wilson line, as appears in, for instance,
ref.~\cite{Gervais:1979fv}.
(Recall that a static heavy quark propagator is represented just
by a straight Wilson line and the $\delta m$ corresponds to
an additive static heavy quark mass shift.
The renormalization pattern (\ref{EQ:wilson-line_renormalization})
is manifest in the picture of $z$-field or static heavy quark.)
The power-law divergence is contained in the $\delta m$
in the exponential factor, leaving only logarithmic divergences in $Z_z$.
For the non-local quark bilinears $O_{\cal C}$, which we are interested in,
the renormalization pattern has also known to be
\begin{eqnarray}
O_{\cal C}=Z_{\psi,z}e^{\delta m \ell({\cal C})}O_{\cal C}^{\rm ren},
\label{EQ:non-local-bilinear_renormalization}
\end{eqnarray}
where the factor $Z_{\psi, z}$ contains the quark and $z$-field field
renormalization as well as the quark--$z$-field vertex renormalization.
Again, $Z_{\psi, z}$ does not include the power divergence
and $\delta m$ in the exponential factor,
which corresponds to that in eq.~(\ref{EQ:wilson-line_renormalization}),
takes all the responsibility for the power divergence.
\footnote{As far as we recognize, the perturbative renormalizability
up to two-loop level has been known by the analogy to the static heavy-light
currents. All-order perturbative renomalizability should be addressed in
the future.
Nonperturbative renormalizability is believed to exist, because the
existence of the continuum limit of the heavy-light system has been
numerically checked (e.g.~ref.~\cite{DellaMorte:2005nwx}).}
Knowing the renormalization pattern of the non-local operator
(\ref{EQ:non-local-bilinear_renormalization}), we wish to define
\begin{eqnarray}
{\cal M}_{\delta z}^{\rm S}(P_z)
 =e^{-\delta m|\delta z|}{\cal M}_{\delta z}(P_z),
\label{EQ:subtracted_matrix_element}
\end{eqnarray}
as a completely power divergence free matrix element,
\footnote{On the lattice, the exponential factor for the liner
divergence subtraction should be $e^{-\ln(1+\delta m)|\delta z|}$.}
and hence we propose ``power divergence subtracted''
quasi quark distribution functions by
\begin{eqnarray}
\widetilde{q}^{\rm S}(\tilde{x}, P_z)
=\int\frac{d\delta z}{2\pi}e^{-i\tilde{x}P_z\delta z}
{\cal M}_{\delta z}^{\rm S}(P_z),
\label{EQ:subtracted_quasi}
\end{eqnarray}
which would make the theory well-defined.

To determine $\delta m$, we follow a strategy introduced
in ref.~\cite{Musch:2010ka}, in which a static $q\bar{q}$ potential is adopted.
The static $q\bar{q}$ potential with finite separation $R$,
$V(R)$, can be defined using an $R\times T$ Wilson loop:
\begin{eqnarray}
 W_{R\times T}\propto e^{-V(R)T}\;\;\; (T\rightarrow {\rm large}).
\end{eqnarray}
By taking into account the renormalization of the Wilson loop:
\begin{eqnarray}
W_{R\times T}=e^{\delta m(2R+2T)+4\nu}W_{R\times T}^{\rm ren},
\end{eqnarray}
where $\nu$ represents a renormalization constant from corners of the
$R\times T$ rectangle,
it is concluded that the renormalization of the potential is written as
\begin{eqnarray}
 V^{\rm ren}(R)=V(R)+2\delta m.
\label{EQ:renormalization_of_V}
\end{eqnarray}
There is a degree of freedom to determine $\delta m$
in eq.~(\ref{EQ:renormalization_of_V}), 
we therefore need to fix $\delta m$ by imposing a renomalization condition:
\begin{eqnarray}
 V^{\rm ren}(R_0)=V_0\longrightarrow
\delta m=\frac{1}{2}\left(V_0-V(R_0)\right).
\label{EQ:condition_for_potential}
\end{eqnarray}
The choice of $R_0$ and, hence, $V_0$
in eq.~(\ref{EQ:condition_for_potential}) is arbitrary (scheme).

In the lattice simulation, the subtraction of the linear divergence should be
nonperturbatively carried out.
In our scheme, the Wilson loop has to be measured to extract the static
potential and set the mass renormalization $\delta m$
by eq.~(\ref{EQ:condition_for_potential}).
In the continuum quasi distribution calculation and also
the perturbative matching between continuum and lattice,
the perturbative expression of $V(R)$ is used.
Its continuum one-loop perturbative expression is known to be:
\begin{eqnarray}
V(R)=-g^2C_F\frac{1}{4\pi R}+g^2C_F\int_{k_{\perp 0}}\frac{1}{k_{\perp 0}^2}
+{\cal O}(g^4),
\end{eqnarray}
where $k_{\perp 0}^2=k_1^2+k_2^2+k_3^2$,
and leads to an one-loop amplitude
\begin{eqnarray}
{\cal M}_{\delta z}^{\rm S}(P_3=0)
&=&
\left[1+\left(\frac{g}{4\pi}\right)^2C_F
\left({\cal A}^{\rm subt}(\delta z)-\frac{2\pi|\delta z|}{R_0}\right)
+O(g^4)\right]
\left[{\cal M}_{\delta z}^{\rm S}(P_3=0)\right]^{\rm tree},
\nonumber\\
\label{EQ:one-loop_subtracted_amplitude}
\end{eqnarray}
in which ${\cal A}^{\rm subt}(\delta z)$ is defined so that a linear
divergence in the tadpole-type diagram is subtracted as
\begin{eqnarray}
I_{\rm tadpole}^{\rm S}(\delta z)=
I_{\rm tadpole}(\delta z)+(4\pi)^2\frac{T_{d-1}}{2}|\delta z|,
\label{EQ:I_3-subt}
\end{eqnarray}
with $d-1$ dimensional tadpole integral
\begin{eqnarray}
T_{d-1}=\int_{k_{\perp z}}\frac{1}{k_{\perp z}^2}.
\label{EQ:tadpole_integral}  
\end{eqnarray}

We make a comment about matching procedure in the momentum space.
Throughout this article, we use the coordinate space, where the matching
is carried out before integrating out $\delta z$
in eq.~(\ref{EQ:quasi-quark-PDFs}).
We especially demonstrate the matching of the quasi distribution
between continuum and lattice in the coordinate space.
However, the matchings between normal and quasi distributions have been
done in the momentum space, where the integration of $\delta z$
in eq.~(\ref{EQ:quasi-quark-PDFs}) is performed first
(refs.~\cite{Ji:2013dva, Xiong:2013bka}).
The power divergence subtraction introduced here could affect the matching
calculation in the momentum space.
While we can choose any renormalization condition for $\delta m$
in eq.~(\ref{EQ:condition_for_potential}), non-zero value of $V_0$
would change the tree-level amplitude.
With the subtraction choosing $V_0\not=0$,
the tree-level does not give a delta function,
which forces the matching calculation to be complicated and quite different
from the usual one.


\section{Perturbative matching between continuum and lattice}
\label{SEC:matching_cont_latt}

In this section, we present a calculation strategy for the perturbative
matching between continuum and lattice.
The actual numerical value of the matching factor is shown
in section~\ref{SEC:lattice_pt}.

\subsection{Matching procedure in coordinate space}

We have two choices for the matching; matching by quasi distributions or
non-local matrix elements:
\begin{eqnarray}
\left[\widetilde{q}(\tilde{x}, P_z)\right]^{\rm cont}
&\longleftrightarrow&
\left[\widetilde{q}(\tilde{x}, P_z)\right]^{\rm latt}
\;\;\;\;\;\;\,\mbox{[~in momentum space~]},\\
\left[{\cal M}_{\delta z}(P_z)\right]^{\rm cont}
&\longleftrightarrow&
\left[{\cal M}_{\delta z}(P_z)\right]^{\rm latt}
\;\;\;\;\;\mbox{[~in coordinate space~]},
\end{eqnarray}
in other words, the matching ``in momentum space'' or ``in coordinate space''.
The first one is naive, in which the matching procedure is similar
to the continuum matching between normal and quasi distributions done
in the original context~\cite{Ji:2013dva}.
In this way $\delta z$ is first integrated out,
thus $z$-component of incoming and outgoing momentum at the non-local
operator are fixed to be $\tilde{x}P_z$.
The matching factor would depend on both $\tilde{x}$ and $P_z$ and
the matching can be written in convolution.
On the other hand, we take the second direction in this paper,
which is rather simple from the point of view of the non-local operator itself.

The matching is carried out at each distance scales $\delta z$ in the
non-local operator, hence the matching factor could depend on $\delta z$.
One might think that the renormalization pattern
of the non-local operator could be convolution type,
because the power-law UV divergence exists and then different length
scales could be mixed up.
However, we already know that the renormalization obeys
eqs.~(\ref{EQ:wilson-line_renormalization}) and
(\ref{EQ:non-local-bilinear_renormalization}),
that is, the renormalization is multiplicative, meaning
there could be no mixings between different length scales.
In summary, the orchestration in the renormalization shown in
eqs.~(\ref{EQ:wilson-line_renormalization}) and
(\ref{EQ:non-local-bilinear_renormalization})
controls the matching pattern to be
\begin{eqnarray}
\left[{\cal M}_{\delta z}^{\rm S}(P_z)\right]^{\rm cont}
=Z(\delta z)
\left[{\cal M}_{\delta z}^{\rm S}(P_z)\right]^{\rm latt},
\end{eqnarray}
where the factor is $P_z$-independent and there is no mixing
between different length of $\delta z$.
Because the matching factor is independent of external momenta
in the coordinate space strategy, we set the external momenta to be zero
in our matching calculation below.

In this paper, we introduce a UV cutoff to regulate UV divergence in
the continuum side,
which is originally used in Ji's paper \cite{Ji:2013dva}.
Although we concentrate on the UV cutoff scheme, the dimensional regularization
can also be used after the power divergence subtraction.
We here mention the dimensionality of the UV cutoff.
As we sit in Euclidean space and the non-local operator is elongated only in
one direction ($z$-direction), it is natural to think the system is decomposed
into $3+1$, thus three dimensional UV cutoff could be straightforward.
In practice, we can easily obtain analytic forms for the one-loop amplitude
using the three dimensional UV cutoff.
While this choice itself is fine to regulate the UV divergence,
two dimensional cutoff is more suitable to match to normal distributions,
where the two dimension is chosen to be perpendicular both to $t$ and
$z$-directions.
Two dimensional UV cutoff was actually introduced
in ref.~\cite{Ji:2013dva} and has been used for the matching.
In our calculation, we can also introduce the two dimensional UV cutoff,
the expressions, however, cannot be written in simple analytic forms
as we see soon.

\subsection{One-loop amplitude in continuum with three dimensional
UV cutoff}

We partly showed the one-loop amplitude for the quasi non-local matrix element
in eq.~(\ref{EQ:one-loop_amplitude}) and its power divergence subtracted one
obtained just by replacing $I_{\rm tadpole}(\delta z)$
with $I_{\rm tadpole}^{\rm S}(\delta z)$ (\ref{EQ:I_3-subt}).
We proceed further calculations using the three dimensional UV cutoff scheme
in this subsection.

We set the UV cutoff in perpendicular direction to $z$, $k_{\perp z}\leq\mu$.
One-loop coefficients for the vertex corrections are analytically obtained:
\begin{eqnarray}
I_{\rm vertex}(\delta z)
&=&
\left.2\left(
{\rm Ei}(-k_{\perp z})-(2+k_{\perp z})e^{-k_{\perp z}}\right)
\right|_{k_{\perp z}=\lambda|\delta z|}^{\mu|\delta z|},\\
I_{\rm sail}(\delta z)
&=&
4\ln\frac{\mu}{\lambda}
+\left.4\left(
-{\rm Ei}(-k_{\perp z})+e^{-k_{\perp z}}\right)
\right|_{k_{\perp z}=\lambda|\delta z|}^{\mu|\delta z|},\\
I_{\rm tadpole}^{\rm S}(\delta z)
&=&
4\ln\frac{\mu}{\lambda}
-\left.4{\rm Ei}(-k_{\perp z})
\right|_{k_{\perp z}=\lambda|\delta z|}^{\mu|\delta z|},
\label{EQ:3dim_I_tad}
\end{eqnarray}
where ${\rm Ei}(x)$ is the exponential integral defined by
\begin{eqnarray}
{\rm Ei}(x)=-\int_{-x}^{\infty}dt\frac{e^{-t}}{t},
\end{eqnarray}
which has asymptotic behavior,
${\rm Ei}(-x)\xrightarrow[x\rightarrow\infty]{}0$ and
${\rm Ei}(-x)\xrightarrow[x\rightarrow0]{}\ln x+\gamma_E$,
where $\gamma_E$ is the Euler-Mascheroni constant.
$\delta z\rightarrow0$ limit:
\begin{eqnarray}
I_{\rm vertex}(\delta z)\xrightarrow[\delta z\rightarrow0]{}
2\ln\frac{\mu}{\lambda},\;\;\;
I_{\rm sail}(\delta z)\xrightarrow[\delta z\rightarrow0]{}0,\;\;\;
I_{\rm tadpole}^{\rm S}(\delta z)\xrightarrow[\delta z\rightarrow0]{}0,
\end{eqnarray}
reproduces the local operator result.
The UV and IR structure of the each diagrams at finite $\delta z$ is:
\begin{eqnarray}
&&I_{\rm vertex}(\delta z)\xrightarrow[\rm UV]{}0,\;\;\;\;\;\;\;\;\;\;\;\;
I_{\rm sail}(\delta z)\xrightarrow[\rm UV]{}\mbox{log div},\;\;\;
I_{\rm tadpole}^{\rm S}(\delta z)\xrightarrow[\rm UV]{}\mbox{log div},
\\
&&I_{\rm vertex}(\delta z)\xrightarrow[\rm IR]{}\mbox{log div},\;\;\;\;
I_{\rm sail}(\delta z)\xrightarrow[\rm IR]{}\mbox{finite},\;\;\;\;\;\;\;
I_{\rm tadpole}^{\rm S}(\delta z)\xrightarrow[\rm IR]{}\mbox{finite},
\end{eqnarray}
where we note that the vertex type diagram is protected
from the UV singularity
by finite quark field separation $\delta z\not=0$.

The wave function part is trivial for the covariant gauge.
The quark self-energy with external momentum $p$ at one-loop
is described by a usual sunset diagram
(figure~\ref{FIG:feynman_diagram_sunset}), which gives
\begin{eqnarray}
\Sigma_{\rm sunset}(p)&=&
\int_kG_{\mu\nu}^{AB}(k)V_{\nu}^B(p,k+p)S_q(k+p)V_{\mu}^A(k+p,p)\nonumber\\
&=&-2ig^2C_F\not{\!p}
\int_0^1d\eta\int_k\frac{1-\eta}{[k^2+\eta(1-\eta)p^2]^2},
\end{eqnarray}
and
\begin{eqnarray}
\left.\frac{\partial\Sigma_{\rm sunset}(p)}{\partial\not{\!p}}\right|_{p=0}
&=&-ig^2C_F\int_k\frac{1}{k^4}.
\end{eqnarray}
By introducing UV and IR regulator in $\perp z$ direction,
we obtain
\begin{eqnarray}
F=-2\ln\frac{\mu}{\lambda}.
\end{eqnarray}

The obtained result in the continuum is summarized
showing total $O(g^2)$ part:
\begin{eqnarray}
{\cal A}^{\rm S}(\delta z)
&=&
2\left(3\ln\frac{\mu}{\lambda}
-3{\rm Ei}(-\mu|\delta z|)+3{\rm Ei}(-\lambda|\delta z|)
-\mu|\delta z|e^{-\mu|\delta z|}
+\lambda|\delta z|e^{-\lambda|\delta z|}
\right).\;\;\;\;\;
\label{EQ:one-loop_coeff_continuum}
\end{eqnarray}
The $\delta z\rightarrow0$ limit again reproduce a local vector current result,
which is zero, because it is a conserved current.

\subsection{One-loop amplitude in continuum with two dimensional
UV cutoff}

We now turn onto the two dimensional UV cutoff case.
The directions for the cutoff are set to be perpendicular to $t$ and $z$,
which we simply call $\perp$ direction.

\begin{figure}
\begin{center}
\includegraphics[scale=0.45, viewport = 0 0 530 460, clip]
{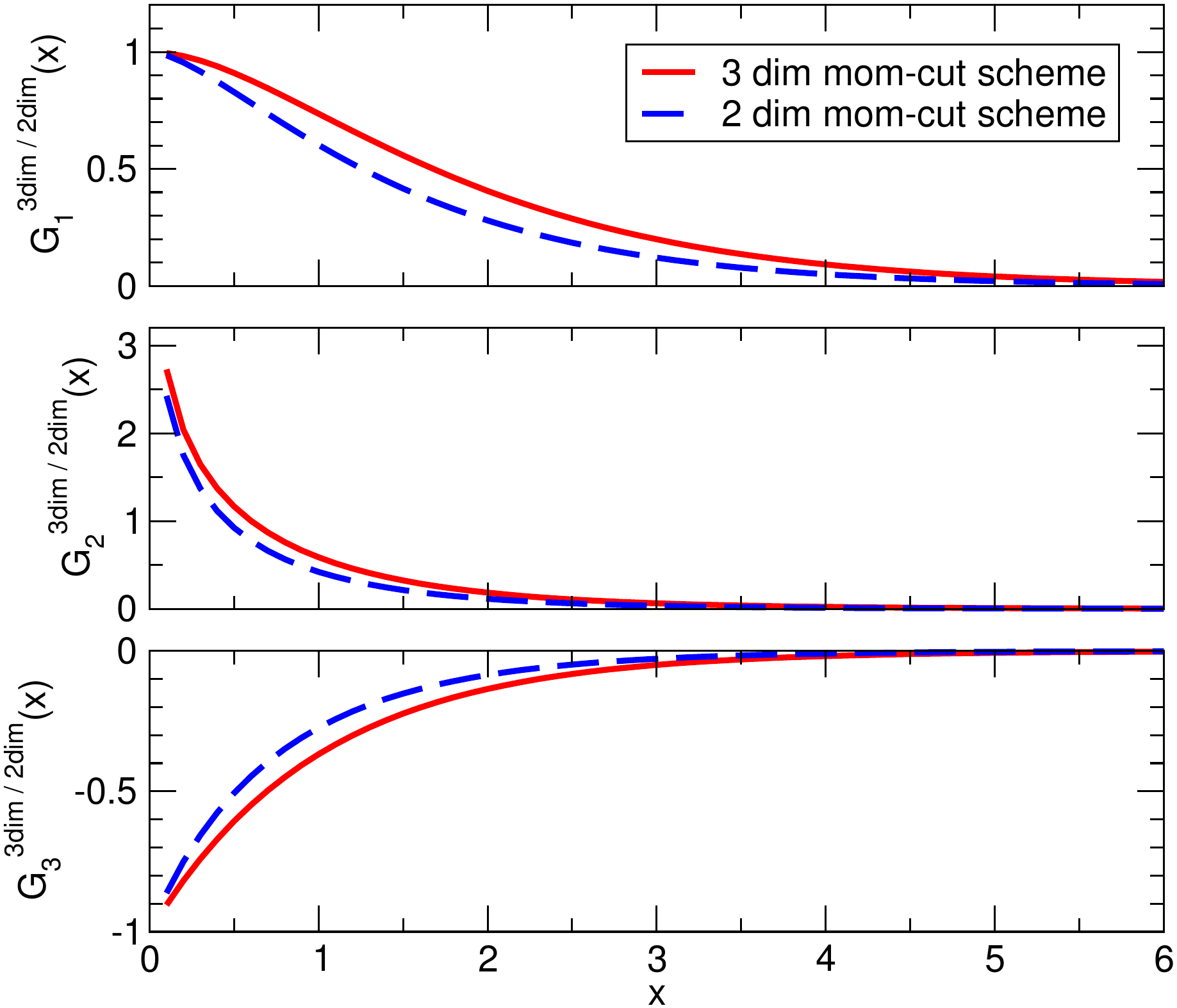}
\caption{Comparison of functions in table~\ref{TAB:correspondence_cutoff}
between two dimensional and three dimensional cutoff scheme.}
\label{FIG:correspondence_3dim-2dim}
\end{center}
\end{figure}

We first integrate out $k_3$ and successively carry out the $k_{\perp}$
integration introducing UV and IR cutoff, $\mu$ and $\lambda$,
then we obtain for $\delta z=0$:
\begin{eqnarray}
I_{\rm vertex}(\delta z=0)=2\ln\frac{\mu}{\lambda},\;\;\;
I_{\rm sail}(\delta z=0)=0,\;\;\;
I_{\rm tadpole}^{\rm S}(\delta z=0)=0,
\end{eqnarray}
and for $\delta z\not=0$:
\begin{eqnarray}
I_{\rm vertex}(\delta z\not=0)&=&
-\int_{-\infty}^{\infty}dk_0
\left.\left(k_{\perp}+\frac{1}{\sqrt{k_0^2+1}}\right)
e^{-\sqrt{k_0^2+1}k_{\perp}}
\right|_{k_{\perp}=\lambda|\delta z|}^{\mu|\delta z|},
\\
I_{\rm sail}(\delta z\not=0)&=&
4\ln\frac{\mu}{\lambda}
+2\int_{-\infty}^{\infty}dk_0
\left.
\frac{e^{-\sqrt{k_0^2+1}k_{\perp}}}{\sqrt{k_0^2+1}}
\right|_{k_{\perp}=\lambda|\delta z|}^{\mu|\delta z|},
\\
I_{\rm tadpole}^{\rm S}(\delta z\not=0)&=&
4\ln\frac{\mu}{\lambda}
+2\int_{-\infty}^{\infty}dk_0
\left.\left(\frac{e^{-\sqrt{k_0^2+1}k_{\perp}}}{\sqrt{k_0^2+1}}
+k_{\perp}{\rm Ei}\left[-\sqrt{k_0^2+1}k_{\perp}\right]\right)
\right|_{k_{\perp}=\lambda|\delta z|}^{\mu|\delta z|}.\nonumber\\
\label{EQ:2dim_I_tad} 
\end{eqnarray}
\begin{table}[t]
\begin{center}
\renewcommand{\arraystretch}{2.5}  
\begin{tabular}{lcl}
\hline\hline
\multicolumn{1}{c}{2 dimensional cutoff} &  &
\multicolumn{1}{c}{3 dimensional cutoff} \\
\hline
$\displaystyle
G^{\rm 2dim}_1(|x|)=
\frac{1}{2}\int_{-\infty}^{\infty}dk_0|x|e^{-\sqrt{k_0^2+1}|x|}$
&$\Longleftrightarrow$&
$\displaystyle
G^{\rm 3dim}_1(|x|)=(|x|+1)e^{-|x|}$\\     
$\displaystyle 
G^{\rm 2dim}_2(|x|)=\frac{1}{2}\int_{-\infty}^{\infty}dk_0
\frac{e^{-\sqrt{k_0^2+1}|x|}}{\sqrt{k_0^2+1}}$
&$\Longleftrightarrow$&
$\displaystyle 
G^{\rm 3dim}_2(|x|)=e^{-|x|}-{\rm Ei}\left[-|x|\right]$\\
$\displaystyle
G^{\rm 2dim}_3(|x|)=
\frac{1}{2}\int_{-\infty}^{\infty}dk_0
|x|{\rm Ei}\left[-\sqrt{k_0^2+1}|x|\right]$
&$\Longleftrightarrow$&
$\displaystyle
G^{\rm 3dim}_3(|x|)=-e^{-|x|}$\vspace*{+2mm}\\
\hline\hline
\end{tabular}
 \caption{Correspondence of functions appear in the one-loop amplitude
 between two and three dimensional cutoff.}
\label{TAB:correspondence_cutoff}
 \end{center}
\end{table}
By comparing with the three dimensional cutoff case,
we find a correspondence of functions presented
in table~\ref{TAB:correspondence_cutoff},
whose functional behavior is shown in figure~\ref{FIG:correspondence_3dim-2dim}.
While they show similar dependences on $x$ between the two cutoff schemes,
slight deviations are observed.

The total $O(g^2)$ part amounts to 
\begin{eqnarray}
{\cal A}^{\rm S}(\delta z)
&=&
2\left(3\ln\frac{\mu}{\lambda}
-G_1^{\rm 2dim}(\mu|\delta z|)+G_1^{\rm 2dim}(\lambda|\delta z|)
+3G_2^{\rm 2dim}(\mu|\delta z|)-3G_2^{\rm 2dim}(\lambda|\delta z|)
\right.
\nonumber\\
&&\left.
+2G_3^{\rm 2dim}(\mu|\delta z|)-2G_3^{\rm 2dim}(\lambda|\delta z|)
\right),
\label{EQ:one-loop_coeff_continuum_2dim}
\end{eqnarray}
which is the two dimensional cutoff counterpart of
eq.~(\ref{EQ:one-loop_coeff_continuum}).

\subsection{Implementation of the one-loop matching
  between continuum and lattice}
\label{SEC:implementation}

In this paper we set a common choice of $V_0=V(R_0)$ between continuum
and lattice in the matching, making the matching factor is irrelevant
to both $V_0$ and $R_0$.
We use UV cutoff scheme, in which the scale is set to be $\mu$.
Instead, we can also use dimensional regularization in the continuum part,
if necessary.

The one-loop perturbative matching is carried out between a scale
$\mu$ in the continuum and $a^{-1}$ in the lattice side,
and can be written in general by
\begin{eqnarray}
Z(\delta z; \mu\leftrightarrow a^{-1})
&=&
1+\left(\frac{g}{4\pi}\right)^2C_F
\left(\left[{\cal A}^{\rm S}(\delta z; \mu)\right]^{\rm cont}
-\left[{\cal A}^{\rm S}(\delta z; a^{-1})\right]^{\rm latt}\right)
+{\cal O}(g^4),\;\;\;
\end{eqnarray}
with power divergence subtracted one-loop coefficients in continuum and
lattice:
\begin{eqnarray}
\left[{\cal A}^{\rm S}(\delta z; \mu)\right]^{\rm cont}
 &=&
(4\pi)^2\int_{-\infty}^{\infty}\frac{d^4k}{(2\pi)^4}
\left[{\cal F}^{\rm S}(\delta z)\right]^{\rm cont}
\nonumber\\
&=&
(4\pi)^2\int_{-\infty}^{\infty}\frac{d^4k}{(2\pi)^4}
\left[{\cal F}(\delta z)\right]^{\rm cont}+(4\pi)^2\frac{T_3}{2}|\delta z|,
\\
\left[{\cal A}^{\rm S}(\delta z; a^{-1})\right]^{\rm latt}
&=&
(4\pi)^2\int_{-\pi/a}^{\pi/a}\frac{d^4k}{(2\pi)^4}
\left[{\cal F}^{\rm S}(\delta z)\right]^{\rm latt}
\nonumber\\
&=&
(4\pi)^2\int_{-\pi/a}^{\pi/a}\frac{d^4k}{(2\pi)^4}
\left[{\cal F}(\delta z)\right]^{\rm latt}
+(4\pi)^2\frac{T_3^{\rm latt}}{2}|\delta z|,
\end{eqnarray}
where $T_3$ is a three dimensional tadpole integral depicted in
eq.~(\ref{EQ:tadpole_integral}), $T_3^{\rm latt}$ is a lattice counterpart
of $T_3$ and
integrals in continuum are assumed to have three dimensional (sphere)
or two dimensional (circle) UV cutoff with a scale $\mu$.

The implementation of the calculation for the matching factor is not
only one way, while the final results should be unique.
We here take a following decomposition for the integrals:
\begin{eqnarray}
\left[{\cal A}^{\rm S}(\delta z; \mu)\right]^{\rm cont}
-\left[{\cal A}^{\rm S}(\delta z; a^{-1})\right]^{\rm latt}
={\cal B}^{(1)}(\delta z)-{\cal B}^{(2)}(\delta z)+{\cal B}^{(3)}(\delta z),
\label{EQ:one-loop_coefficient_matching}
\end{eqnarray}
with definitions for three dimensional cutoff:
\begin{eqnarray}
{\cal B}^{(1)}(\delta z)
&=&
(4\pi)^2\int_{-\infty}^{\infty}\frac{d^4k}{(2\pi)^4}
\left[{\cal F}^{\rm S}(\delta z)\right]^{\rm cont}
\theta(\mu^2-k_0^2-k_1^2-k_2^2)
\theta(k_0^2+k_1^2+k_2^2-\lambda^2),\;\;\;\;\;\;\;\;
\\
{\cal B}^{(2)}(\delta z)
&=&
(4\pi)^2\int_{-\pi/a}^{\pi/a}\frac{d^4k}{(2\pi)^4}
\left\{\left[{\cal F}^{\rm S}(\delta z)\right]^{\rm latt}
-\left[{\cal F}^{\rm S}(\delta z)\right]^{\rm cont}
\theta(\lambda^2-k_0^2-k_1^2-k_2^2)\right\},
\\
{\cal B}^{(3)}(\delta z)
&=&
(4\pi)^2\int_{-\infty}^{\infty}\frac{d^4k}{(2\pi)^4}
\left[{\cal F}^{\rm S}(\delta z)\right]^{\rm cont}
\theta(\lambda^2-k_0^2-k_1^2-k_2^2)
\theta(k_3^2-(\pi/a)^2),
\end{eqnarray}
and for two dimensional cutoff:
\begin{eqnarray}
{\cal B}^{(1)}(\delta z)
&=&
(4\pi)^2\int_{-\infty}^{\infty}\frac{d^4k}{(2\pi)^4}
\left[{\cal F}^{\rm S}(\delta z)\right]^{\rm cont}\theta(\mu^2-k_1^2-k_2^2)
\theta(k_1^2+k_2^2-\lambda^2),
\\
{\cal B}^{(2)}(\delta z)
&=&
(4\pi)^2\int_{-\pi/a}^{\pi/a}\frac{d^4k}{(2\pi)^4}
\left\{\left[{\cal F}^{\rm S}(\delta z)\right]^{\rm latt}
-\left[{\cal F}^{\rm S}(\delta z)\right]^{\rm cont}
\theta(\lambda^2-k_1^2-k_2^2)\right\},
\\
{\cal B}^{(3)}(\delta z)
&=&
(4\pi)^2\int_{-\infty}^{\infty}\frac{d^4k}{(2\pi)^4}
\left[{\cal F}^{\rm S}(\delta z)\right]^{\rm cont}\theta(\lambda^2-k_1^2-k_2^2)
\nonumber\\
&&\hspace*{+47mm}\times
\theta(k_0^2-(\pi/a)^2)\theta(k_3^2-(\pi/a)^2).
\end{eqnarray}
In the above expression, ${\cal B}^{(1)}(\delta z)$ is obtained
analytically in eq.~(\ref{EQ:one-loop_coeff_continuum}) for three dimensional
cutoff and numerically in eq.~(\ref{EQ:one-loop_coeff_continuum_2dim})
for two dimensional cutoff.
${\cal B}^{(2)}(\delta z)$ and ${\cal B}^{(3)}(\delta z)$
need numerical integrations.
Note that the left-hand side of eq.~(\ref{EQ:one-loop_coefficient_matching})
is completely independent of $\lambda$, that is,
the $\lambda$ dependence in the right-hand side
should be canceled out between ${\cal B}^{(1)}(\delta z)$,
${\cal B}^{(2)}(\delta z)$ and ${\cal B}^{(3)}(\delta z)$.


\section{Lattice perturbation and results of the matching factor --
 na\"{\i}ve fermion case}
\label{SEC:lattice_pt}

In this section, we demonstrate calculations of the matching factor
using the lattice perturbation theory.
We follow the implementation described in subsection~\ref{SEC:implementation}
and employ the naive fermion action for lattice fermions
and the standard plaquette action for lattice gluons.
The naive fermion, of course, has a doubling problem and thus is
not practical.
We, however, use it for simplifying the discussion.
The extension to more practical lattice fermions,
such as Wilson fermion and domain-wall fermion,
is trivial task, but just increases a level of complication.

\subsection{Mean-field improvement}

To make the convergence of the lattice perturbation better,
the mean-field (MF) improvement~\cite{Lepage:1992xa} is often employed.
Starting point of the MF improvement is replacing the gluon links
$U_{\mu}$ with $U_{\mu}/u_0$ where $u_0$ denotes MF value of $U_{\mu}$.
In this paper, we choose the fourth-root of the expectation value of
the plaquette $P$ as the definition of $u_0$.
The perturbative expansion of $u_0$ is written as
\begin{eqnarray}
u_0=P^{1/4}=1-g^2C_F\frac{T_{\rm MF}}{2}+{\cal O}(g^4),
\end{eqnarray}
where the MF factor $T_{\rm MF}=1/8$ for the plaquette gluon action.
The MF improvement can be carried out by replacements:
\begin{eqnarray}
\frac{1}{g^2}&\longrightarrow&\frac{P}{g^2},\\
\frac{1}{g_{\overline{\rm MS}}^2}&\longrightarrow&
\frac{1}{g_{\overline{\rm MS}}^2}+2C_FT_{\rm MF},\\
\psi&\longrightarrow&u_0^{1/2}\psi,\\
f^{\rm latt}&\longrightarrow&f^{\rm latt}-(4\pi)^2\frac{T_{\rm MF}}{2},
\end{eqnarray}
where $g_{\overline{\rm MS}}$ is the continuum $\overline{\rm MS}$ coupling.
One might think more replacements are required because the non-local
operator contains a Wilson line with finite length.
They are, however, not needed in our power divergence subtraction scheme
as the effects from the MF improvement are completely canceled out.

\subsection{Link smearing for the Wilson line}

Link smearing is a widely used technique in lattice QCD simulations
to reduce noise and also power divergences contained in the Wilson line.
In this paper, we consider a 3-step hyper-cubic blocking type smearing
~\cite{Hasenfratz:2001hp}, especially parameter choices,
HYP1~\cite{Hasenfratz:2001hp} and HYP2~\cite{DellaMorte:2005yc}.
The smearing changes the Feynman rules and
we use a procedure described in refs.~\cite{Lee:2002fj, DeGrand:2002va}
to include the smearing into the one-loop perturbation,
in which original gluon fields $A_{\mu}(x)$ are replaced by smeared ones as
\begin{eqnarray}
A_{\mu}(x)\longrightarrow B_{\mu}(x)=\sum_{\nu}h_{\mu\nu}(x)A_{\nu}(x),
\end{eqnarray}
where the function $h_{\mu\nu}(x)$ depends on the smearing,
and this leads to a modification to Feynman rules for
the non-local operator (Appendix~\ref{APPENDIX:gluon_link_smearing}).
In general the link smearing could change the MF improvement procedure.
However, we do not need the change after subtracting the power divergence,
if the smearing is carried out only on the Wilson line
in the non-local operator.

\subsection{Numerical results of the matching coefficient at one-loop level} 

The Feynman rules for the lattice perturbation and the expressions
of integrals for the numerical evaluations are presented in appendices.
In this subsection, we show the numerical values of the matching
coefficients.

We use UV cutoff scheme for the continuum, whose cutoff scale is
set to be $\mu=a^{-1}$.
When we subtract the linear UV divergence, the effects, which depends on the
choice of subtraction condition (\ref{EQ:condition_for_potential}),
come into eq.~(\ref{EQ:one-loop_subtracted_amplitude}) and
their counter parts in the lattice side.
However, the terms are canceled out in the matching between continuum
and lattice, we thus do not need to care about the choice of the subtraction
in this matching.
\begin{figure}
\begin{center}
\parbox{48mm}{
\begin{center}
\includegraphics[scale=0.62, viewport = 0 0 230 480, clip]
{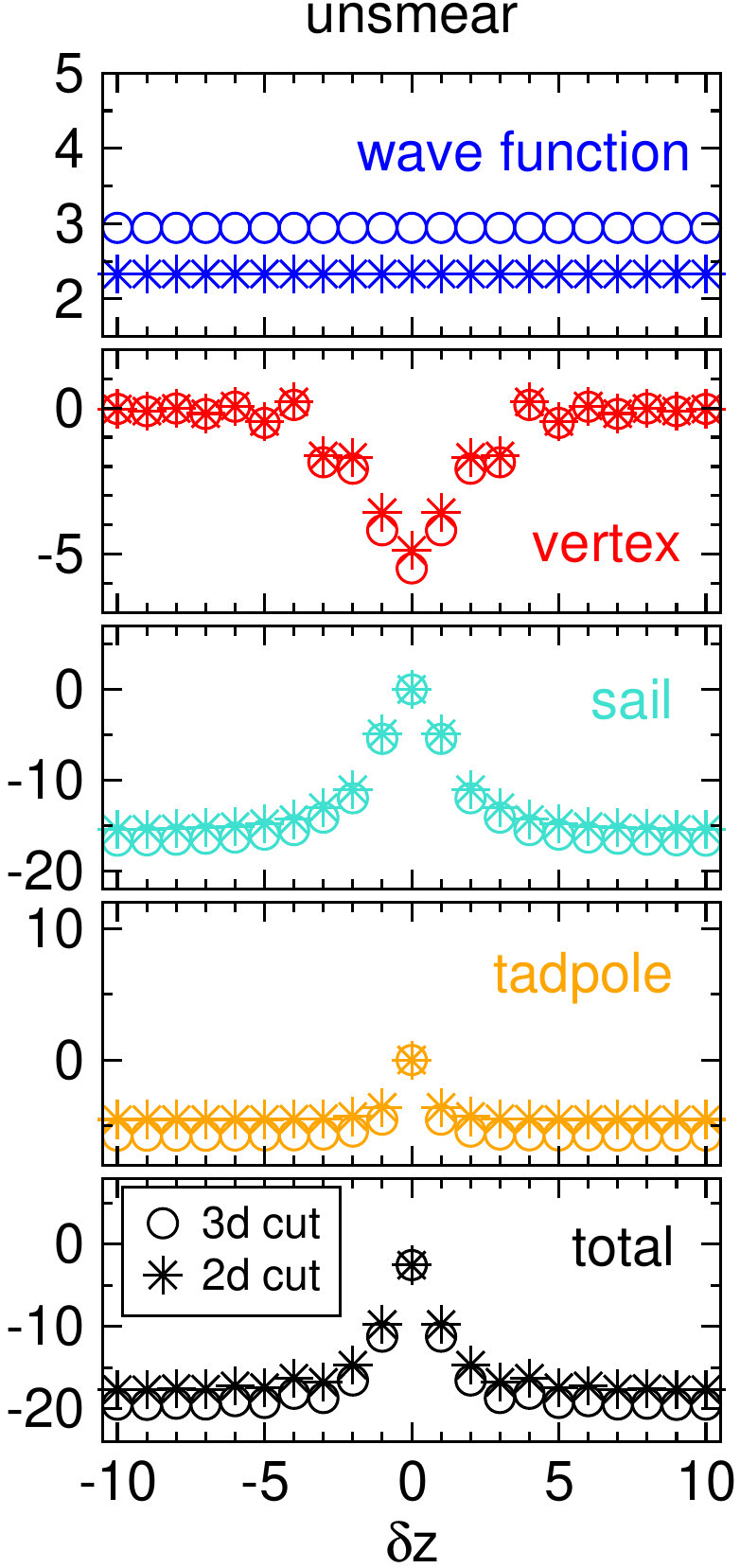}
\end{center}
}
\parbox{48mm}{
\begin{center}
\includegraphics[scale=0.62, viewport = 0 0 230 480, clip]
{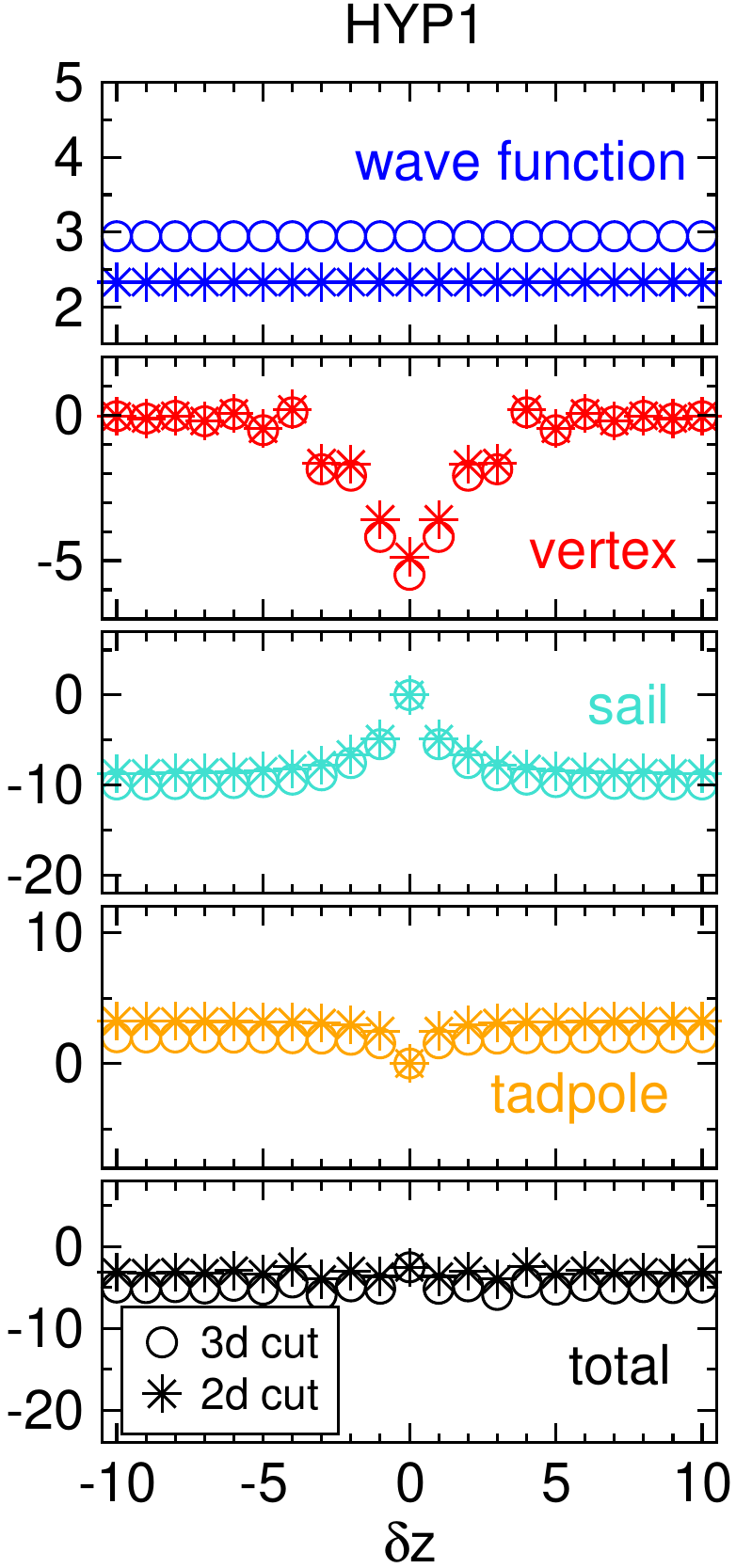}
\end{center}
}
\parbox{48mm}{
\begin{center} 
\includegraphics[scale=0.62, viewport = 0 0 230 480, clip]
{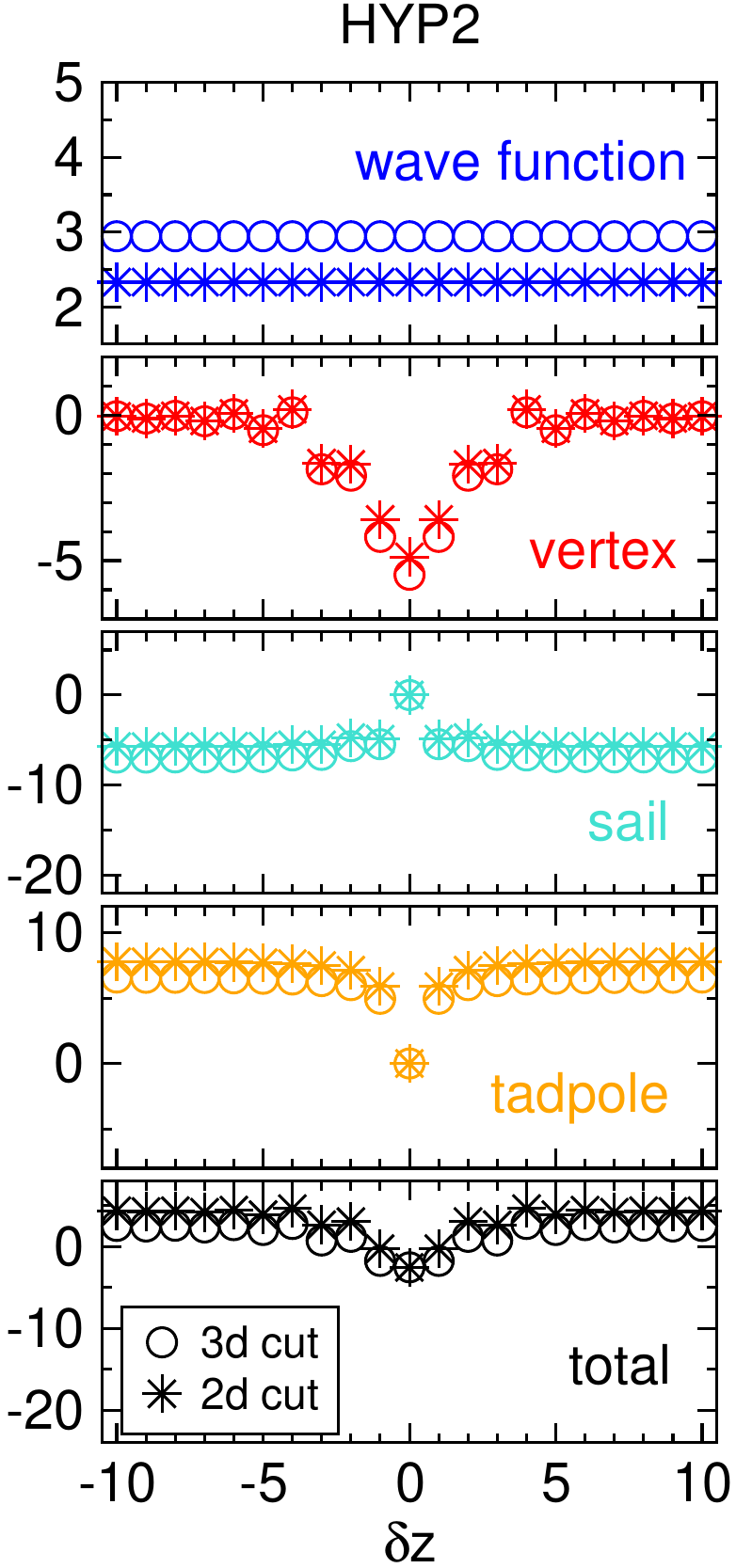} 
\end{center}
}
 \caption{One-loop matching coefficients for each individual diagrams:
 quark self-energy, vertex-type, sail-type and tadpole-type,
 as well as their total contribution.
 The linear divergence is subtracted and the MF improvement is used.
 Three cases of gluon link smearing are considered for a Wilson line in the
 non-local operator: unsmear (left),
 HYP1~\cite{Hasenfratz:2001hp} (center) and
 HYP2~\cite{DellaMorte:2005yc} (right).
 Both three dimensional (circle symbols) and two dimensional
 (star symbols) UV cutoff
 cases are shown.}
 \label{FIG:one-loop_matching_ coefficients} 
 \end{center}
\end{figure}
\begin{figure}
\begin{center}
\parbox{51mm}{
\begin{center}
\includegraphics[scale=0.62, viewport = 0 0 230 420, clip]
{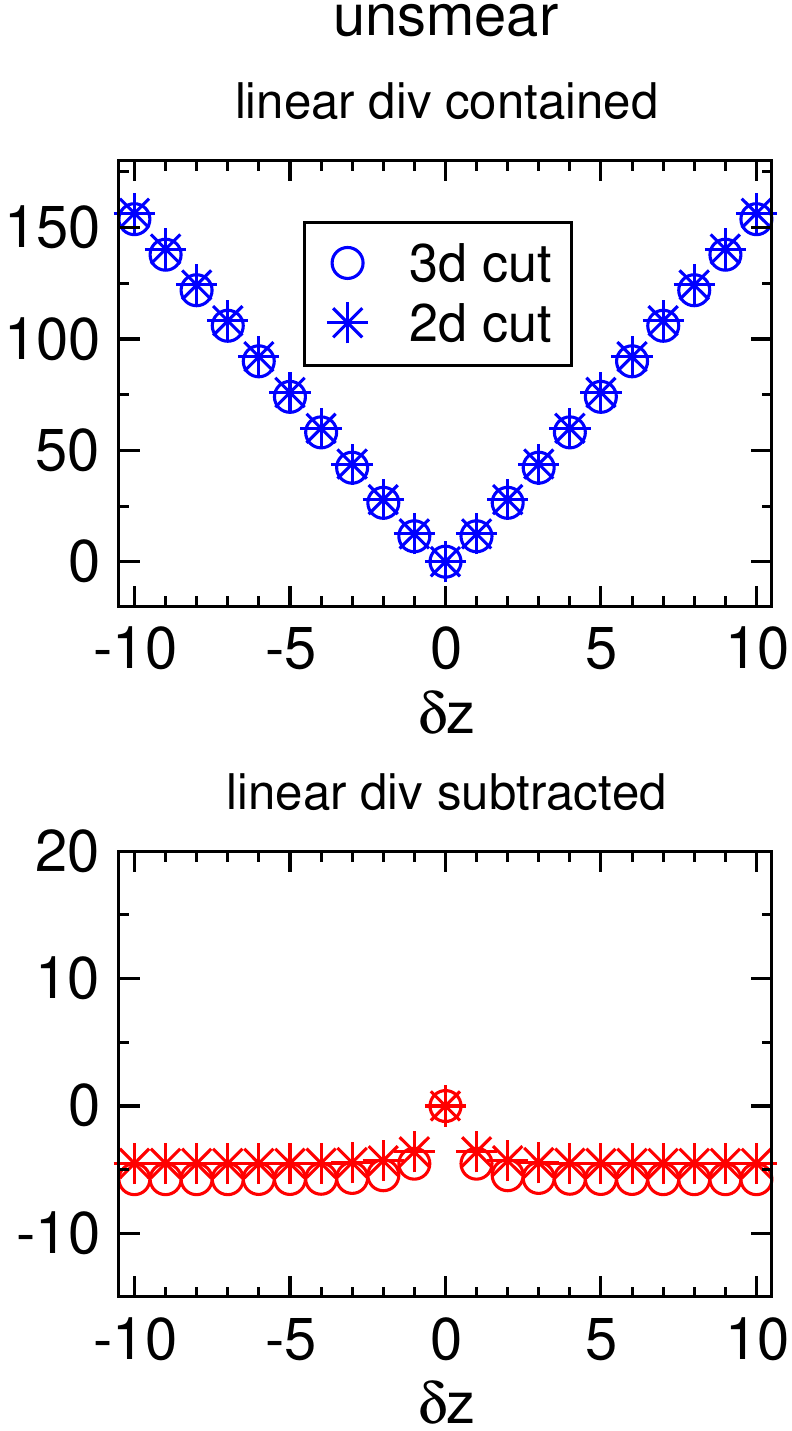}
\end{center}
}
\parbox{48mm}{
\begin{center}
\includegraphics[scale=0.62, viewport = 0 0 230 420, clip]
{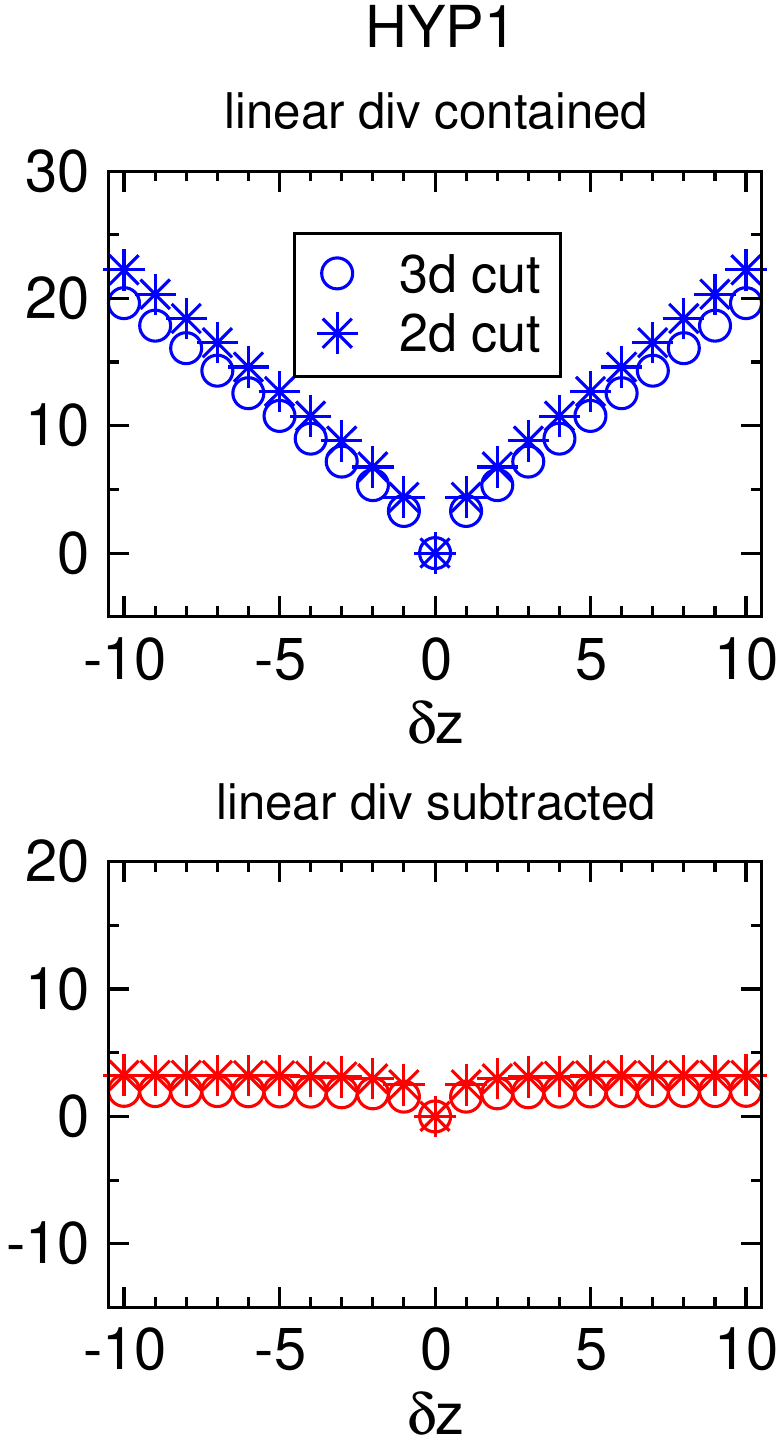}
\end{center}
}
\parbox{48mm}{
\begin{center} 
\includegraphics[scale=0.62, viewport = 0 0 230 420, clip]
{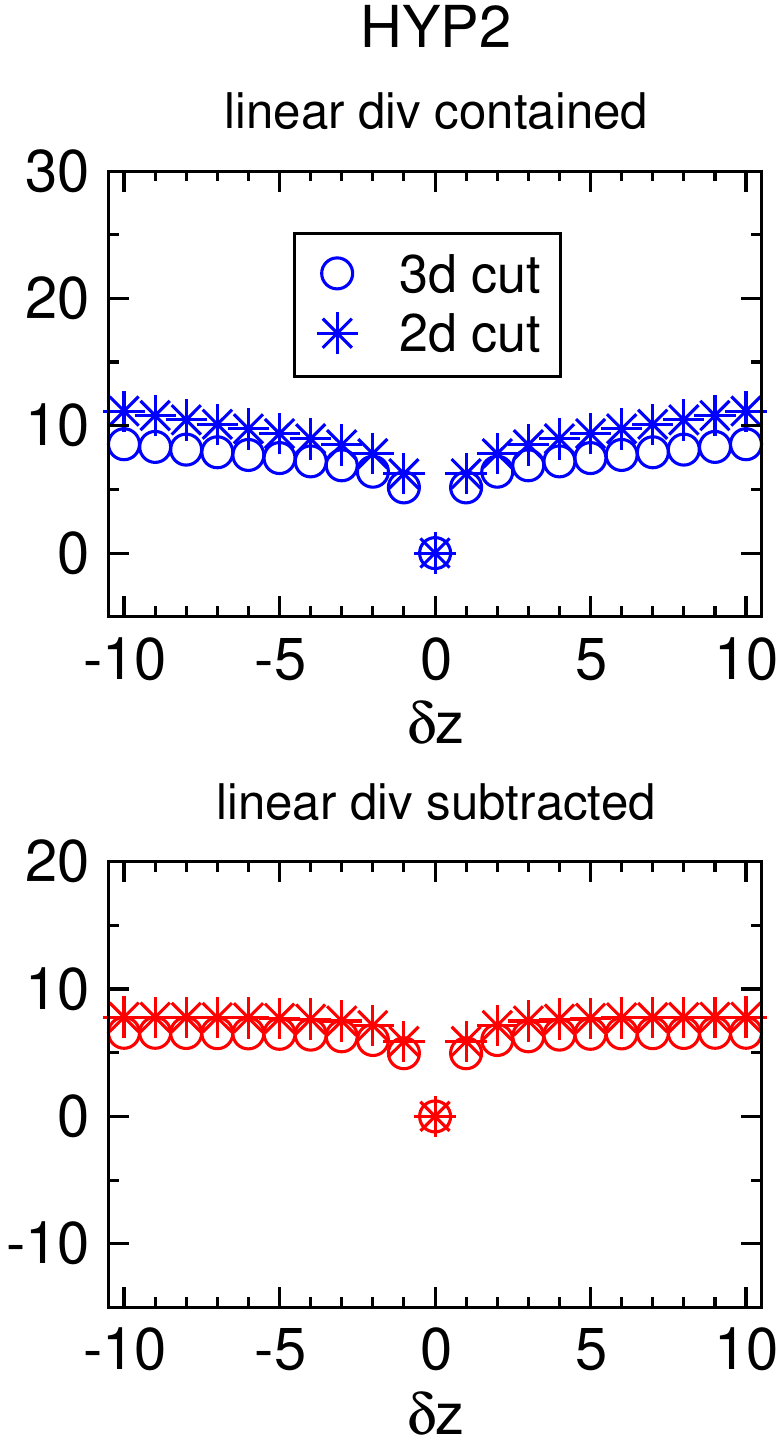} 
\end{center}
}
\caption{Tadpole-type diagram contributions to one-loop matching coefficients.
 The linear divergence is contained in the upper figures,
 while it is subtracted in the bottom figures.
 MF improvement is not used here.}
 \label{FIG:one-loop_ matching_ coefficients_tadpole} 
 \end{center}
\end{figure}

The numerical value of one-loop matching coefficients is presented in
figure~\ref{FIG:one-loop_matching_ coefficients},
where contributions from each individual diagrams,
quark self-energy (figure~\ref{FIG:feynman_diagram_sunset}),
vertex-type (figure~\ref{FIG:feynman_diagram_deltaGamma}, left),
sail-type (figure~\ref{FIG:feynman_diagram_deltaGamma}, middle two) and
tadpole-type (figure~\ref{FIG:feynman_diagram_deltaGamma}, right),
are separately shown, as well as their total coefficients.
The linear UV divergences are subtracted, and the MF improvement is used,
which only affects to the value for the wave function.
The one-loop contributions from the wave function and the vertex-type
do not depend on link smearings on the Wilson line (unsmear, HYP1 and HYP2),
whilst the sail-type and tadpole-type do.
We put comments on the results:
\begin{itemize}
 \item The $\delta z$ dependent behavior
       is restricted in the small $\delta z$ region.
       This is because the difference of the continuum and the lattice
       sits only on UV structure.
       (For the tadpole-type, this notion is valid only after the power
       divergence is subtracted.)
 \item For the vertex-type, the value is non-zero at $\delta z=0$,
       because we use the local vector current on the lattice.
 \item The vertex-type contribution at large $\delta z$ becomes zero,
       because long gluon lines are suppressed.
       For the sail- and tadpole-type, contributions remain finite at
       the large $\delta z$, where the difference in the short distance
       between continuum and lattice is seen.
 \item The unsmeared case shows the same sign between sail- and tadpole-type,
       resulting in huge value of total one-loop coefficients at larger
       $\delta z$, while the smeared cases give these contributions
       opposite sign, which leads a good cancellation.
       The link smearing would be important from the view of
       perturbative accuracy in the matching.
\end{itemize}

Finally, we would like to mention the effectiveness of the link smearing
on especially tadpole-type contribution and power divergence subtraction.
Figure~\ref{FIG:one-loop_ matching_ coefficients_tadpole} shows
the tadpole-type contribution with (lower figures) and without (upper figures)
the linear divergence subtraction.
In the upper figures, the linearly increasing behavior in $\delta z$ is
observed due to existence of the power divergence,
showing that the perturbation does not make sense at all
at large $\delta z$.
\footnote{While it would be better to implement the MF improvement for
the tadpole-type contribution, 
we here do not use it believing this does not lose
any essence of the discussion.}
After the divergence subtraction, the linearly increasing behavior
is completely removed (lower figures).
If we do not smear the Wilson line (left figure), the power divergence
is so intense, however the link smearings make the divergence mild.
Especially, by using HYP2 smearing, the power divergence vanishes
in large part before the subtraction.


\section{Summary and outlook}
\label{SEC:summary}

The lattice QCD calculation of the parton distribution functions
on the lattice had been impossible, however, recent proposals
~\cite{Ji:2013dva, Ma:2014jla} opened a possibility to enable us
to overcome the difficulty.
So far, our reach in this method is not at quantitative level and
more refinements are necessary for this approach giving us
valuable information to our understanding for the nature of nuclear physics.
The improvements are demanded in both continuum and lattice side.
For the continuum side, more rigid matching procedure between
normal and quasi distributions is significant.
We proposed the power divergence subtraction for making the matching
well-defined.
In this paper, we concentrated on the quasi parton distributions,
which was introduced Ji's original paper~\cite{Ji:2013dva} using
the large momentum effective theory.
We would also be able to introduce other type of observables to extract
our target distribution functions by the context of the collinear
factorization approach~\cite{Ma:2014jla},
where the new measurable observables could be designed
to have more simplicity and better accessibility.
For the lattice side, more rigorous treatment is required toward more
quantitative information, such as matching to the continuum,
which we demonstrated perturbatively in this paper.
While nonperturbative matching would be preferable in the end,
this perturbative work gives a good baseline for those future study.
Although the perturbative matching factor was calculated only for impractical
na\"{\i}ve lattice fermion action in this paper,
the extension to the practical lattice actions would be trivial.
Another technical challenge is accessibility to large hadron momenta
on the lattice, where large statistical errors could ruin the accuracy of
the numerical simulation, while the large hadron momenta make the
perturbative error in the hard part small.
There have recently been some proposals to this issue, such as
refs.~\cite{Roberts:2012tp, DellaMorte:2012xc, Bali:2016lva},
in which quark field smearings suitable to the high hadron momenta are
discussed.

Apparently, our quest is not only to parton distribution functions,
but also to transverse-momentum dependent parton densities (TMDs) and
generalized parton distributions (GPDs) for the three dimensional full scan
of nucleon structure.
Future experiments such as EIC focus on the search for the full image
of the structure to reveal the nucleon spin,
and again the lattice QCD could provide complementary valuable information.
The extension of the method we discussed here to the TMDs and GPDs needs
more refinements, it is, however, essential to explore the three dimensional
structure of the nucleon.


\acknowledgments

The authors benefited from discussions with Taku Izubuchi and
Andreas Sch\"{a}fer.
S. Y. is supported by JSPS Strategic Young Researcher Overseas Visits
Program for Accelerating Brain Circulation (No.R2411).




\appendix

\section{Feynman rules for lattice perturbation}
\label{SEC:feynman_rules}

In this appendix, we present our definitions for the lattice
perturbative calculation.
We mostly follow definitions in ref.~\cite{Capitani:2002mp}.

\subsection{Fourier transform of fields and some definitions}

\begin{eqnarray}
\psi(x)=\int_{-\pi/a}^{\pi/a}\frac{d^dk}{(2\pi)^d}e^{ik\cdot x}\psi(k),\;\;\;
\psi(k)=a^d\sum_xe^{-ik\cdot x}\psi(x),
\end{eqnarray}
\begin{eqnarray}
\overline{\psi}(x)=
\int_{-\pi/a}^{\pi/a}\frac{d^dk}{(2\pi)^d}e^{-ik\cdot x}\overline{\psi}(k),
\;\;\;
\overline{\psi}(k)=a^d\sum_xe^{ik\cdot x}\overline{\psi}(x),
\end{eqnarray}
\begin{eqnarray}
A_{\mu}(x)=\int_{-\pi/a}^{\pi/a}\frac{d^dk}{(2\pi)^d}e^{ik\cdot x}A_{\mu}(k),
\;\;\;
A_{\mu}(k)=a^d\sum_xe^{-ik\cdot x}A_{\mu}(x), 
\end{eqnarray}
\begin{eqnarray}
\delta_{xy}=a^d\int_{-\pi/a}^{\pi/a}\frac{d^dk}{(2\pi)^d}e^{ik\cdot (x-y)},
\;\;\;
\delta^{(d)}(k)=\frac{a^d}{(2\pi)^d}\sum_xe^{-ik\cdot x},
\end{eqnarray}
\begin{eqnarray}
U_{\mu}(x)=e^{igaA_{\mu}(x+\hat{\mu}/2)},\;\;\;
U_{\mu}^{\dagger}(x)=e^{-igaA_{\mu}(x+\hat{\mu}/2)},
\end{eqnarray}
\begin{eqnarray}
\hat{k}_{\mu}=2\sin\frac{ak_{\mu}}{2},\;\;\;
\not{\!\hat{k}}=\sum_{\mu}\gamma_{\mu}\hat{k}_{\mu},\;\;\;
\hat{k}^2=\sum_{\mu}\hat{k}_{\mu}^2.
\end{eqnarray}

\subsection{Gluon propagator
            (Plaquette with Covariant gauge ($\partial_{\mu}A_{\mu}=0$))}

\begin{eqnarray}
G_{\mu\nu}^{AB}(k)=
\delta^{AB}\frac{a^2}{\hat{k}^2}\left(\delta_{\mu\nu}
-(1-\alpha)\frac{\hat{k}_{\mu}\hat{k}_{\nu}}{\hat{k}^2}\right)
\xrightarrow[a\rightarrow0]{}
\delta^{AB}\frac{1}{k^2}
\left(\delta_{\mu\nu}-(1-\alpha)\frac{k_{\mu}k_{\nu}}{k^2}\right),
\end{eqnarray}
where $\alpha$ is a gauge fixing parameter.
In this paper, we take $\alpha=1$, Feynman gauge.

\subsection{Quark propagator (Naive fermion)}

Quark propagator with quark mass $m$ using naive fermion can be written as:
\begin{eqnarray}
S_q(k)=
a\frac{-i\sum_{\mu}\gamma_{\mu}\sin ak_{\mu}+am}
{\sum_{\mu}\sin^2ak_{\mu}+(am)^2}
\xrightarrow[a\rightarrow0]{}
\frac{-i\not{\!k}+m}{k^2+m^2}.
\end{eqnarray}
This fermion possesses the chiral symmetry, while has doublers.
In this paper, the naive fermion is employed just for simplicity.

\subsection{Quark-gluon vertex (Naive fermion)}

\subsubsection*{Quark-quark-gluon:}

\begin{eqnarray}
V^A_{\mu}(p, q)=-gT^A
i\gamma_{\mu}\cos\frac{a(p+q)_{\mu}}{2}
\xrightarrow[a\rightarrow0]{}
-igT^A\gamma_{\mu}.
\end{eqnarray}

\subsubsection*{Quark-quark-gluon-gluon:}

\begin{eqnarray}
V^{AB}_{\mu\nu}(p, q)
=\frac{1}{2}ag^2\delta_{\mu\nu}
\left\{T^A, T^B\right\}
i\gamma_{\mu}\sin\frac{a(p+q)_{\mu}}{2}
\xrightarrow[a\rightarrow0]{}0.
\end{eqnarray}

\subsection{Non-local quark bilinear operator (covariant gauge)}
\label{APPENDIX:feynman_rules_non-local_operator}

In this appendix we provide a derivation of the Feynman rule for the
quasi quark PDF non-local operator defined by eq.~(\ref{EQ:non-local_operator}).
We here do not assume $A_3=0$ gauge, which gives trivial rule for the
non-local operator.
In the derivation, we first consider the lattice discretization,
and then take continuum limit.

The lattice discretized version of the Wilson line operator is
\begin{eqnarray}
U_3(x+\hat{\bm 3}|\delta z|; x)
&=&
U_3^{\dagger}(x+\hat{\bm 3}(N-1)a)
U_3^{\dagger}(x+\hat{\bm 3}(N-2)a)\cdots
U_3^{\dagger}(x+\hat{\bm 3}a)U_3^{\dagger}(x),
\\
U_3(x-\hat{\bm 3}|\delta z|; x)
&=&
U_3(x-\hat{\bm 3}Na))
U_3(x-\hat{\bm 3}(N-1)a)\cdots
U_3(x-\hat{\bm 3}2a)U_3(x-\hat{\bm 3}a),\;\;\;\;\;\;
\end{eqnarray}
where we define $|\delta z|=aN$.
By expanding the gauge link in $A_3(x)$,
\begin{eqnarray}
U_3(x\pm\hat{\bm 3}|\delta z|; x)
&=&1
\mp iga\sum_{n=0}^{N-1}
A_3\left(x\pm\hat{\bm 3}\left(n+\frac{1}{2}\right)a\right)
-\frac{(ga)^2}{2}\sum_{n=0}^{N-1}
A_3\left(x\pm\hat{\bm 3}\left(n+\frac{1}{2}\right)a\right)^2\nonumber\\
&&-(ga)^2\sum_{m=0}^{N-2}\sum_{n=m+1}^{N-1}
A_3\left(x\pm\hat{\bm 3}\left(n+\frac{1}{2}\right)a\right)
A_3\left(x\pm\hat{\bm 3}\left(m+\frac{1}{2}\right)a\right)
+{\cal O}(g^3)
\nonumber\\
&=&1-iga\int_{-\pi/a}^{\pi/a}\frac{d^4k}{(2\pi)^4}
A_3(k)e^{ik\cdot x}\frac{e^{\pm ik_3aN}-1}{i\hat{k}_3}
\nonumber\\
&&
\mp\frac{(ga)^2}{2}\int_{-\pi/a}^{\pi/a}\frac{d^4k}{(2\pi)^4}
\int_{-\pi/a}^{\pi/a}\frac{d^4l}{(2\pi)^4}A_3(k)A_3(l)e^{i(k+l)\cdot x}
\frac{e^{\pm i(k_3+l_3)aN}-1}{i\widehat{(k_3+l_3)}}
\nonumber\\
&&
-(ga)^2\int_{-\pi/a}^{\pi/a}\frac{d^4k}{(2\pi)^4}
\int_{-\pi/a}^{\pi/a}\frac{d^4l}{(2\pi)^4}A_3(k)A_3(l)
e^{i(k+l)\cdot x}e^{\pm i\frac{(k_3+l_3)a}{2}}
\frac{e^{\pm ik_3a}}{e^{\pm ik_3a}-1}
\nonumber\\
&&
\times
\left(\frac{e^{\pm ik_3a(N-1)}-e^{\pm i(k_3+l_3)a(N-1)}}{1-e^{\pm il_3a}}
-\frac{1-e^{\pm i(k_3+l_3)a(N-1)}}{1-e^{\pm i(k_3+l_3)a}}\right)
+{\cal O}(g^3),
\label{EQ:derivation_nonlocal}
\end{eqnarray}
we obtain rules for the non-local operator on lattice and then continuum:
\begin{eqnarray}
O_{\delta z}^{(0)}(p, q)&=&\gamma_3\delta(p-q)e^{-ip_3\delta z}
\nonumber\\
&\xrightarrow[a\rightarrow0]{}&
\gamma_3\delta(p-q)e^{-ip_3\delta z},
\\
O_{\delta z}^{(1)\mu, A}(p, q, k)&=&
igaT^A\gamma_3\delta^{\mu 3}\delta(p-q-k)e^{-ip_3\delta z}
\frac{1-e^{ik_3\delta z}}{i\hat{k}_3}
\nonumber\\
&\xrightarrow[a\rightarrow0]{}&
igaT^A\gamma_3\delta^{\mu 3}\delta(p-q-k)e^{-ip_3\delta z}
\frac{1-e^{ik_3\delta z}}{ik_3},\\
O_{\delta z}^{(2)\mu\nu, AB}(p, q, k)&=&
-g^2\{T^A,T^B\}\gamma_3\delta^{\mu 3}\delta^{\nu 3}\delta(p-q)e^{-ip_3\delta z}
\left(a^2\frac{1-e^{ik_3\delta z}}{\hat{k}_3^2}
-a\frac{\delta z}{i\hat{k}_3}e^{\frac{\delta z}{|\delta z|}i\frac{k_3a}{2}}
\right)
\nonumber\\
&&-\frac{g^2a}{2}\{T^A,T^B\}\gamma_3\delta^{\mu 3}\delta^{\nu 3}
\delta(p-q)e^{-ip_3\delta z}|\delta z|
\nonumber\\
&\xrightarrow[a\rightarrow0]{}{}&
-g^2\{T^A,T^B\}\gamma_3\delta^{\mu 3}\delta^{\nu 3}\delta(p-q)e^{-ip_3\delta z}
\left(\frac{1-e^{ik_3\delta z}}{k_3^2}-\frac{\delta z}{ik_3}\right),
\end{eqnarray}
where $O_{\delta z}^{(2)\mu\nu, AB}(p, q, k)$ is obtained by setting $k+l=0$ in
eq.~(\ref{EQ:derivation_nonlocal}),
which is enough for the one-loop calculation.

\subsection{Gluon link smearing}
\label{APPENDIX:gluon_link_smearing}

For the Wilson line in the non-local operator, we use link smearing.
This smearing changes the Feynman rules~\cite{Lee:2002fj,DeGrand:2002va}.
In this case the rules of the non-local operator are modified as
\begin{eqnarray}
O_{\delta z}^{(1)\mu, A}(p, q, k)&\longrightarrow&
\tilde{h}_{\mu\mu'}(k)O_{\delta z}^{(1)\mu', A}(p, q, k),\\
O_{\delta z}^{(2)\mu\nu, AB}(p, q, k)&\longrightarrow&
\tilde{h}_{\mu\mu'}(k)O_{\delta z}^{(2)\mu'\nu', AB}(p, q, k)
\tilde{h}_{\nu'\nu}(k),
\end{eqnarray}
where
\begin{eqnarray}
\tilde{h}_{\mu\nu}(k)
&=&\delta_{\mu\nu}
\left[1-\frac{\alpha_1}{6}\sum_{\rho}\hat{k}_{\rho}^2\Omega_{\mu\rho}(k)\right]
+\frac{\alpha_1}{6}\hat{k}_{\mu}\hat{k}_{\nu}\Omega_{\mu\nu}(k),\\
\Omega_{\mu\nu}(k)&=&1+\alpha_2(1+\alpha_3)
-\frac{\alpha_2}{4}(1+2\alpha_3)(\hat{k}^2-\hat{k}_{\mu}^2-\hat{k}_{\nu}^2)
+\frac{\alpha_2\alpha_3}{4}\prod_{\eta\not=\mu,\nu}\hat{k}_{\eta}^2,
\end{eqnarray}
with HYP smearing parameters $(\alpha_1, \alpha_2, \alpha_3)$.
We use parameter choices:
\begin{eqnarray}
(\alpha_1, \alpha_2, \alpha_3)=
\begin{cases}
(0.75, 0.6, 0.3) & \mbox{:~HYP1~\cite{Hasenfratz:2001hp}} \\
(1.0, 1.0, 0.5)  & \mbox{:~HYP2~\cite{DellaMorte:2005yc}}.  
\end{cases}
\end{eqnarray}


\section{One-loop correction on the lattice}
\label{SEC:one-loop_lattice}

We show the one-loop expression of the vertex correction and the quark
self-energy obtained by the lattice perturbation.

\subsection{Vertex correction}

For the vertex correction, one-loop coefficients for each diagrams are
expressed as
\begin{eqnarray}
I_{\rm vertex}^{\rm latt}(\delta z)&=&
(4\pi)^2\int_{-\pi/a}^{\pi/a}\frac{d^4k}{(2\pi)^4}
{\cal F}_{\rm vertex}^{\rm latt},\\
I_{\rm sail}^{\rm latt}(\delta z)&=&
(4\pi)^2\int_{-\pi/a}^{\pi/a}\frac{d^4k}{(2\pi)^4}
{\cal F}_{\rm sail}^{\rm latt},\\
I_{\rm tadpole}^{\rm latt}(\delta z)&=&
(4\pi)^2\int_{-\pi/a}^{\pi/a}\frac{d^4k}{(2\pi)^4}
{\cal F}_{\rm tadpole}^{\rm latt},
\end{eqnarray}
where
\begin{eqnarray}
{\cal F}_{\rm vertex}^{\rm latt}
&=&
\frac{a^4}{\hat{k}^2}\frac{1}{(\sum_{\mu}\sin^2(ak_{\mu}))^2}
\left\{\sum_{\mu}\sin^2(ak_{\mu})-2\sin^2(ak_3)\right\}
\\
&&\times
\left\{\sum_{\mu}\cos^2\left(\frac{ak_{\mu}}{2}\right)
-2\cos^2\left(\frac{ak_3}{2}\right)\right\}
\cos(k_3\delta z)
\xrightarrow[ak\ll 1]{}
2\left(\frac{1}{k^4}-\frac{2k_3^2}{k^6}\right)\cos(k_3\delta z),
\nonumber\\
{\cal F}_{\rm sail}^{\rm latt}
&=&
\frac{2a^4}{\hat{k}^2}
\frac{1}{\sum_{\mu}\sin^2(ak_{\mu})}
\cos^2\left(\frac{ak_3}{2}\right)
\left(1-\cos(k_3\delta z)\right)
\xrightarrow[ak\ll 1]{}
\frac{2}{k^4}(1-\cos(k_3\delta z)),
\\
{\cal F}_{\rm tadpole}^{\rm latt}
&=&
\frac{a^4}{\hat{k}^2}\frac{\cos(k_3\delta z)-1}{\hat{k}_3^2}
\xrightarrow[ak\ll 1]{}
\frac{1}{k^2}\frac{\cos(k_3\delta z)-1}{k_3^2}.
\end{eqnarray}

\subsection{Wave function}

 \begin{figure}
\begin{center}
\parbox{45mm}{
\begin{center}
\includegraphics[scale=0.3, viewport = 0 0 320 250, clip]
{./Figures/Feynman/sunset.pdf}
$\Sigma_{\rm sunset}(p)$
\end{center}
}
\parbox{45mm}{ 
\begin{center}
\includegraphics[scale=0.3, viewport = 0 0 320 240, clip]
{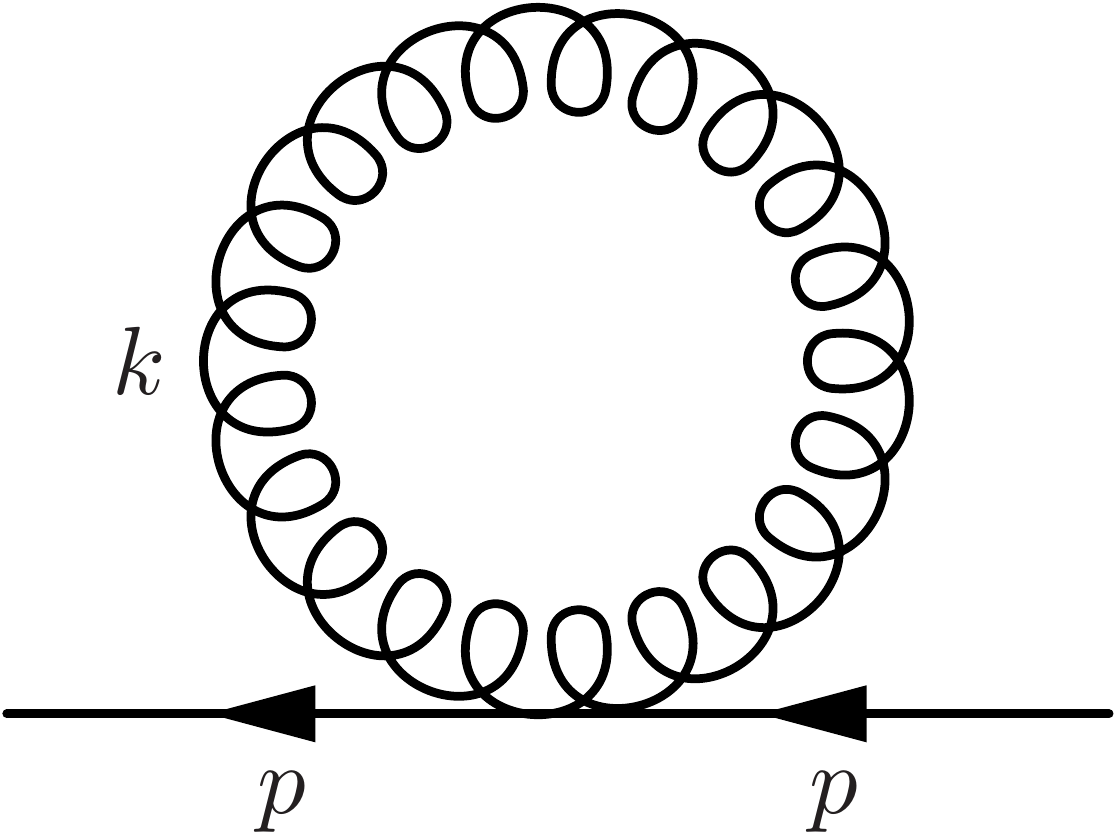}
$\Sigma_{\rm tad}(p)$
\end{center}
}
 \caption{Sunset (left) and tadpole (right) diagram
 for quark wave-function renormalization on the lattice.}
\label{FIG:feynman_diagram_sunset_tadpole_lattice}
 \end{center}  
\end{figure}
The sunset and tadpole diagram
(figure.~\ref{FIG:feynman_diagram_sunset_tadpole_lattice})
produce one-loop correction
to quark self-energy $\Sigma_{\rm sunset}^{\rm latt}(p)$
and $\Sigma_{\rm tad}^{\rm latt}(p)$,
respectively, and give one-loop coefficient of the wave function:
\begin{eqnarray}
F^{\rm latt}
&=&
\left(\left(\frac{g}{4\pi}\right)^2C_F\right)^{-1}
\left(
\left.
\frac{\partial \Sigma_{\rm sunset}^{\rm latt}(p)}{\partial i\!\!\not\!p}\right|_{p=0}
+\left.
\frac{\partial \Sigma_{\rm tad}^{\rm latt}(p)}{\partial i\!\!\not\!p}\right|_{p=0}
\right)
\nonumber\\
&=&
(4\pi)^2\int_{-\pi/a}^{\pi/a}\frac{d^4k}{(2\pi)^4}
{\cal F}_{\Sigma_{\rm sunset}}^{\rm latt}
+(4\pi)^2\int_{-\pi/a}^{\pi/a}\frac{d^4k}{(2\pi)^4}
{\cal F}_{\Sigma_{\rm tad}}^{\rm latt},
\end{eqnarray}
where
\begin{eqnarray}
{\cal F}_{\Sigma_{\rm sunset}}^{\rm latt}&=&
-\frac{1}{4}\frac{a^4}{\hat{k}^2}
\frac{\sum_{\mu}\cos(ak_{\mu})}{(\sum_{\nu}\sin^2(ak_{\nu}))^2}
\nonumber\\
&&\times
\left(2\sin^2(ak_{\mu})-\sum_{\nu}\sin^2(ak_{\nu})\right)
\left(2\cos^2\left(\frac{ak_{\mu}}{2}\right)
-\sum_{\nu}\cos^2\left(\frac{ak_{\nu}}{2}\right)\right)
-\frac{a^4}{4}\frac{1}{\hat{k}^2}
\nonumber\\
&\xrightarrow[ak\ll 1]{}&
-\frac{1}{k^4}
\\
{\cal F}_{\Sigma_{\rm tad}}^{\rm latt}&=&
\frac{a^4}{2}\frac{1}{\hat{k}^2}
\xrightarrow[ak\ll 1]{}
\frac{a^2}{2}\frac{1}{k^2}.
\end{eqnarray}
An integral included in the expressions above gives a numerical value:
\begin{eqnarray}
T_4^{\rm latt}&=&
\int_{-\pi/a}^{\pi/a}\frac{d^4k}{(2\pi)^4}\frac{a^4}{\hat{k}^2}=0.154933,
\end{eqnarray}
where IR divergences are absent.




\end{document}